\documentclass[aps,preprintnumbers,eqsecnum,amsmath,amssymb,nofootinbib]{revtex4}
\usepackage{graphicx}
\usepackage{dcolumn}
\usepackage{bm}
\usepackage{epsfig}

\begin{document}
\title{Heavy-to-light form factors: sum rules on the light cone and beyond}  
\author{Wolfgang Lucha$^{a}$, Dmitri Melikhov$^{a,b}$ and Silvano Simula$^{c}$}
\affiliation{
$^a$ Institute for High Energy Physics, Austrian Academy of Sciences, Nikolsdorfergasse 18, A-1050, Vienna, Austria\\
$^b$ Nuclear Physics Institute, Moscow State University, 119992, Moscow, Russia\\
$^c$ INFN, Sezione di Roma III, Via della Vasca Navale 84, I-00146, Roma, Italy}
\date{\today}
\begin{abstract}
We report the first systematic analysis of the off-light-cone effects in 
sum rules for heavy-to-light form factors. These effects are investigated in a model
based on scalar constituents, which allows a technically rather simple
analysis but has the essential features of the analogous QCD calculation.  
The correlator relevant for the extraction of the heavy-to-light form factor 
is calculated in two different ways: 
first, by adopting the full Bethe--Salpeter amplitude of the light meson   
and, second, by performing the expansion of this amplitude near the light cone $x^2=0$. 
We demonstrate that the contributions to the correlator from the light-cone term $x^2=0$ and the
off-light-cone terms $x^2\ne 0$ have the same order in the $1/m_Q$ expansion.  
The light-cone correlator, corresponding to $x^2=0$, is shown to systematically 
overestimate the full correlator, the difference being $\sim \Lambda_{\rm QCD}/\delta$, 
with $\delta$ the continuum subtraction parameter of order 1 GeV. 
Numerically, this difference is found to be 
$10\div 20$\%.  
\end{abstract}
\maketitle
\vspace{-.5cm}
\section{Introduction}
QCD sum rules on the light cone \cite{lcsr} have been extensively applied to various exclusive 
form factors, in particular, to weak heavy-to-light transition form factors 
(see, e.g., \cite{ck}). 
Within this method, the relevant correlators are obtained as power
expansions in $x^2$ near the 
light cone (LC) $x^2=0$ in terms of the distribution amplitudes of increasing twist. 
In practice, however, only few lowest-twist distribution amplitudes are known with a 
reasonable accuracy. 
Thus one has to rely on calculations which take into account only lowest powers of
$x^2$, at most up to terms linear in $x^2$, without a systematic study 
of the effects related to higher powers of $x^2$ (\cite{ck,bb,bz} and references therein). 
However, the off-LC $(x^2\ne 0)$ contributions to the correlator are not parametrically 
suppressed compared to the contribution evaluated at $x^2=0$:  
for instance, Braun and Halperin \cite{bh} studied the pion elastic form 
factor with light-cone sum rules and found that contributions 
to the correlator from terms 
corresponding to higher powers of $x^2$ in the pion Bethe--Salpeter (BS) amplitude 
$\langle 0|T\bar u(x)\gamma_\mu\gamma_5 d(0)|\pi(p)\rangle$ 
have in general the same $\sim 1/q^4$ behaviour.  
As we show here, a similar situation occurs for the heavy-to-light form factor: 
the contributions to the correlator coming from 
the LC and the off-LC terms in the BS amplitude of the light meson  
have the same dependence in $1/m_Q$, $m_Q$ being the heavy-quark mass. 
Therefore, relying on calculations performed at $x^2=0$ does not seem to be safe 
and the effects of all powers of $x^2$ should be summed and taken into account in the sum rule. 

In this paper, we study the correlator relevant for the extraction of the 
heavy-to-light form factors with sum rules and propose a method which allows one to obtain and 
take into account the full $x^2$-dependence of the BS amplitude of the light meson.  
For the sake of argument, we present the analysis for the model with scalar constituents,  
the heavy scalar $Q$ of the mass $m_Q$ and the light scalar $\varphi$ of the mass $m$  
(which we name ``quarks'' throughout the paper). We study the weak transition of a heavy spinless 
bound state $M_Q(Q\varphi)$ to the light spinless bound state $M(\varphi\varphi)$ (which we refer to as 
mesons) induced by the weak heavy-to-light $Q\to\varphi$ quark transition. 
The analysis of this model is technically simpler but at the same time 
allows one to study some essential features of the corresponding QCD case. 

The paper is organized as follows: 

In Sec.~\ref{sect_bs}, we consider the BS amplitude of the light meson 
$\Psi_{\rm BS}(x,p)=\langle 0|T\varphi(x)\varphi(0)|M(p)\rangle$ and its expansion in powers of $x^2$.    
In Sec.~\ref{sect_ff}, we study the correlator 
$i\int d^4x\exp(ipx) \langle 0|T \varphi(x)Q(x)Q(0)\varphi(0)|M(p')\rangle$, 
relevant for the extraction of the 
$M_Q\to M$ form factor within the method of sum rules. First, we show that the use 
of the Nakanishi representation for
the BS amplitude of the light meson amounts to a 
summation of all the off-LC effects. We refer to the correlator obtained by this procedure 
as the full correlator. Second, we make use of the 
expansion of the BS amplitude near the LC and obtain a different form
of the correlator corresponding to this LC expansion.  
We show that for the quark--quark interaction dominated by 
one-boson exchange at small distances (similar to QCD, where the
interaction is dominated by one-gluon exchange) any term corresponding to $(x^2)^n$ 
in the BS amplitude gives a contribution to the correlator which behaves as $1/m_Q^2$,  
independent of $n$. Keeping the term $x^2=0$ only, gives the LC correlator. 
We then discuss the procedure of making the Borelization of the correlator and obtaining the sum rule for the full and the LC
correlators. 
Section~\ref{numerical} presents the numerical analysis of the full and the LC correlators for 
charm and beauty decays. It is shown that at $q^2\simeq 0$ the LC correlator systematically overestimates the full 
correlator, the difference being numerically $10\div 20$\%, independent of the mass of the heavy quark. 
Section~\ref{conclusions} summarizes our results.  
  
\section{\label{sect_bs}The Bethe--Salpeter amplitude and its expansion near the light cone}
The BS amplitude is defined according to
\begin{eqnarray}
\label{BS}
\Psi_{\rm BS}(x,p')=\langle 0|T\varphi(x)\varphi(0)|M(p')\rangle=\Psi(x^2,xp',p'^2=M^2). 
\end{eqnarray}
As a function of the variable $xp'$, the amplitude may be represented by the Fourier integral 
\begin{eqnarray}
\label{nak1}
\Psi_{\rm BS}(x,p')=
\int\limits_0^1 d\xi \exp({-i\xi p'x})K(x^2,\xi),  
\end{eqnarray}
where the $\xi$-integration runs from $0$ to $1$ as follows from the analytic properties of 
Feynman diagrams. Nakanishi proposed to parametrize the kernel
$K(x^2,\xi)$ as \cite{nakanishi}
\begin{eqnarray}
\label{nak2}
K(x^2,\xi)=\frac{1}{(2\pi)^4 i}\int\limits_0^\infty dz\, G(z,\xi)
\int d^4k' 
\frac{\exp({-ik'x})}{[\,z+m^2-\xi(1-\xi)M^2-k'^2-i0]^3},   
\end{eqnarray} 
where the function $G(z,\xi)$ has no singularities in the integration regions in $z$ and $\xi$. 
The BS equation for the function $\Psi_{\rm BS}$ leads to an equation for
the function $G(z,\xi)$. 

The $k'$-integral in (\ref{nak2}) is the second derivative w.r.t.\ $\mu^2$ of the 
Feynman propagator of a scalar particle of mass $\mu$ 
[with $\mu^2=z+m^2-\xi(1-\xi)M^2$] in coordinate space. Making use of the explicit expression for 
this propagator \cite{bogolyubov}, we obtain the following expansion near the light cone $x^2=0$: 
\begin{eqnarray}
\label{prop}
\frac{1}{(2\pi)^4 i}\int  d^4k'\exp({-ik'x})
\frac{1}{(\mu^2-k'^2-i0)^3}&=&
\frac12\frac{\partial^2}{(\partial \mu^2)^2}
\left\{\frac{1}{(2\pi)^4 i}\int  d^4k'\exp({-ik'x})
\frac{1}{\mu^2-k'^2-i0}\right\}
\nonumber\\
&=&\frac{1}{32\pi^2 \mu^2}+\frac{x^2}{128\pi^2}\left[1-2\gamma_E-\log\left({-\mu^2 x^2}/{4}\right)\right]+
O(x^4). 
\end{eqnarray}
Interestingly, this is not a pure power series but it contains also terms involving $\log(-x^2)$. 
Inserting (\ref{prop}) into (\ref{nak2}) leads to the light-cone expansion of $K(x^2,\xi)$:  
\begin{eqnarray}
\label{logseries}
K(x^2,\xi)=g_0(\xi)&+&x^2 \left[\,g_1(\xi)+\log(-x^2 m^2)h_1(\xi)\right]
+x^4  \left[\,g_2(\xi)+\log(-x^2 m^2)h_2(\xi)\right] +\cdots. 
\end{eqnarray}
The functions $g_n$ and $h_n$ here may be expressed in terms of $G(z,\xi)$. 
For instance, 
\begin{eqnarray}
\label{2.6}
g_0(\xi)&=&\frac{1}{32\pi^2}\int\limits_0^\infty dz\, G(z,\xi) \frac{1}{z+m^2-\xi(1-\xi)M^2},
\nonumber\\
g_1(\xi)&=&\frac{1}{128\pi^2}\int\limits_0^\infty dz\, G(z,\xi) 
\left[\,1-2\gamma_E-\log\left(\frac{z+m^2-\xi(1-\xi)M^2}{4m^2}\right)\right],
\nonumber\\
h_1(\xi)&=&-\frac{1}{128\pi^2}\int\limits_0^\infty dz\, G(z,\xi).
\end{eqnarray}
The presence of the logarithmic terms leads to complications:  
Let us go back to Eq.~(\ref{nak2}) and expand the exponential 
$\exp({-ik'x})$. Only the first term corresponding to $x^2=0$ (the light-cone) is finite, whereas all
higher terms lead to divergent $k'$-integrals; this is a consequence of the presence of 
the terms $\log(-x^2)$. After performing the Wick rotation in the $k'$-space, we may cut 
the $k'_E$-integration at $k_E'^2=\Lambda^2$. 
In this way, we obtain a regularized pure {\it power} series in $x^2$ for $K(x^2,\xi)$ 
and thus for $\Psi_{\rm BS}(x,p')$.  
The introduction of the cutoff $\Lambda$ in the momentum-space integrals is equivalent to the
introduction of a regulator $\lambda$ in coordinate space: 
\begin{eqnarray}
\log(-x^2 m^2)\to \log(\lambda-x^2 m^2). 
\end{eqnarray}
The initial expansion (\ref{logseries}) is reproduced by setting $\lambda\to 0$. 
For a nonzero $\lambda$, we can represent $\log(\lambda-x^2 m^2)$ as a power series in 
$x^2$ and insert this expansion in (\ref{logseries}), obtaining 
\begin{eqnarray}
K(x^2,\xi)=\phi_0(\xi)+x^2 \phi_1(\xi,\lambda)+x^4 \phi_2(\xi,\lambda)+\cdots,
\end{eqnarray}
with the first three distribution amplitudes $\phi_n$ given by
\begin{eqnarray}
\label{2.9}
\phi_0(\xi)&=& g_0(\xi), \nonumber\\
\phi_1(\xi,\lambda)&=& g_1(\xi)+\log(\lambda)h_1(\xi), \nonumber\\
\phi_2(\xi,\lambda)&=& g_2(\xi)+\log(\lambda)h_2(\xi)-\frac{1}{\lambda}m^2 h_1(\xi). 
\end{eqnarray}
As a result, instead of the series expansion for the 
BS wave function in terms of powers and logarithms of $x^2$ with finite coefficients, we obtained 
a pure power series, in which only the first term, corresponding to $x^2=0$, is finite, 
whereas all higher distribution amplitudes depend on the regulator parameter $\lambda$ and 
become infinite in the limit $\lambda\to 0$ ($\Lambda\to\infty$).
The regularised LC expansion of the BS amplitude reads 
\begin{eqnarray}
\label{lcexpansion}
\Psi_{\rm BS}(x,p')=\sum_{n=0}^{\infty}(x^2)^n\int\limits_0^1
d\xi \exp({-i p'x\xi})\phi_n(\xi,\lambda). 
\end{eqnarray}
Notice that in our case the full BS amplitude does not depend on the
regulator parameter $\lambda$. 
However, any truncation of the series at some $n$ leads to an explicit
dependence of $\Psi_{\rm BS}(x,p)$ on $\lambda$ in
such a way, that it diverges as $\lambda\to 0$. 

The BS amplitude in momentum space reads
\begin{eqnarray}
\label{bsmom}
\Psi_{\rm BS}(k,p')&=&\int d^4x \exp({ikx})\Psi_{\rm BS}(x,p')=
\frac{1}{i}\int\limits_0^1 d\xi \int\limits_0^\infty dz\,
\frac{G(z,\xi)}{[\,z+m^2-\xi(1-\xi)M^2-(k-\xi p')^2-i0]^3}. 
\end{eqnarray}
The kernel $G(z,\xi)$ determines now the bound-state properties. For instance, 
the decay constant of a bound state $M$ is given by the relation 
\begin{eqnarray}
f_M=\langle 0|\varphi(0)\varphi(0)|M(p')\rangle= \Psi_{\rm BS}(x=0,p').
\end{eqnarray}
Performing the $k'$-integration in Eq.~(\ref{nak2}) for $x=0$ leads to 
\begin{eqnarray}
f_M=\frac1{32\pi^2}
\int\limits_0^1 d\xi \int\limits_0^\infty dz
\frac{G(z,\xi)}{z+m^2-\xi(1-\xi)M^2}.
\end{eqnarray}
The light-cone wave function is related to the kernel $G(z,\xi)$ as follows 
\cite{karmanov}: 
\begin{eqnarray}
\label{2.14}
\Psi_{\rm LC}(\xi,k_\perp)=\xi(1-\xi)\int\limits_0^\infty dz\frac{G(z,\xi)}
{\left[\,z+k_\perp^2+m^2-\xi(1-\xi)M^2\right]^2}.
\end{eqnarray}
Respectively, the distribution amplitudes $\phi_n$ may be calculated as $k_\perp^2$-moments of 
$\Psi_{\rm LC}(\xi,k_\perp)$. 
For the function (\ref{2.14}), only the zero $k_\perp^2$-moment leading to $\phi_0$ is
finite, all higher moments require a cut-off in $k_\perp^2$ precisely as in (\ref{2.9}).

For the case of an interaction between the constituents via exchange of a massless boson, 
the solution to the BS equation in the ladder
approximation takes a simple form \cite{karmanov}: 
\begin{eqnarray}
G(z,\xi)=\delta(z)G(\xi), \qquad G(\xi)=\xi(1-\xi)f(\xi). 
\end{eqnarray}
The function $f(\xi)$ here is a smooth function which takes finite nonzero values at 
the end-points. Only the end-point behavior is crucial 
for the heavy-to-light correlator, therefore the precise form of the function $f(\xi)$ and its normalization 
are not essential for our analysis; of importance for us is only the fact that 
$f(\xi=0)\ne 0$. Hereafter we just set $G(\xi)=m^2\xi(1-\xi)$ and do not care about 
the overall factor $\propto f(0)$. This factor turns out to be the same in the full and the light-cone 
correlators which we calculate and compare in the next section. 

Closing this section, let us emphasize that the $\delta$-functional $z$-dependence of the kernel $G(z,\xi)$ 
leads to a nontrivial $k_\perp^2$-dependence of the light-cone wave function 
\begin{eqnarray}
\label{2.16}
\Psi_{\rm LC}(\xi,k_\perp)=
\frac{\xi(1-\xi)G(\xi)}{\left[\,k_\perp^2+m^2-\xi(1-\xi)M^2\right]^2}.
\end{eqnarray}
The $1/k_\perp^4$ tail of the bound-state LC wave function is a consequence of the 
massless-boson exchange (and is similar in this respect to the power-like behavior of the pion 
light-cone wave function in QCD). 

\section{\label{sect_ff}
Heavy-to-light correlator and the decay form factor}
In this section, we consider the correlator  
\begin{eqnarray}
\Gamma(p^2,q^2)=i \int d^4x \exp({ipx})\langle 0|T \varphi(x)Q(x) Q(0)\varphi(0)|M(p')\rangle  
\end{eqnarray} 
and its Borel transform in the variable $p^2$ relevant for the 
extraction of the $M_Q\to M$ form factor. 
Since $p'^2=M^2$ is fixed, this correlator depends on the two variables $p^2$ and $q^2$. 

\subsection{Borel sum rule for $\Gamma(p^2,q^2)$}
Making use of the hadronic degrees of freedom, by inserting the complete  
system of hadron states into the correlator, we obtain the so-called phenomenological 
representation for 
$\Gamma(p^2,q^2)$, which we denote by $\Gamma_{\rm phen}(p^2,q^2)$: 
\begin{eqnarray}
\label{phen}
\Gamma_{\rm phen}(p^2,q^2)&=&
\frac{1}{(2\pi)^4}\int d^4x \exp({ipx})\langle 0|\varphi(x)Q(x)|M_Q(\tilde p)\rangle
\frac{d^4\tilde p}{M_Q^2-\tilde p^2-i0} 
\langle M_Q(\tilde p)|Q(0)\varphi(0)|M(p')\rangle+
{\rm hadr.\, cont.}
\nonumber\\
&=&F_{M_Q\to M}(q^2)\frac{1}{M_Q^2-p^2-i0}f_{M_Q}+\int\limits_{s_{\rm cont}}^\infty \frac{ds}{s-p^2-i0}
\Delta_{\rm phen}(s,q^2), 
\end{eqnarray}
where $s_{\rm cont}$ is the threshold of the hadronic continuum, containing the 
$M_Q$-meson plus light mesons, 
i.e.,  
$M_Q M$, $M_QMM$, etc.\ states. Respectively, $s_{\rm cont}=(M_Q+M)^2$. 
The decay constant $f_{M_Q}$ of the $M_Q$-meson is defined as
\begin{eqnarray}
f_{M_Q}=\langle 0|T \varphi(0)Q(0)|M_Q(p)\rangle,   
\end{eqnarray}
and the form factor is given by 
\begin{eqnarray}
F_{M_Q\to M}(q^2)=\langle M_Q(p)|Q(0)\varphi(0)|M(p')\rangle, \qquad q\equiv p-p'. 
\end{eqnarray}
Eq.~(\ref{phen}) gives the single dispersion
representation for $\Gamma(p^2,q^2)$ in 
terms of the hadron degrees of freedom. 

One can calculate the correlator $\Gamma(p^2,q^2)$ also by making
use of the underlying field-theoretic degrees of freedom and obtain in this way a different --- 
theoretical --- expression for $\Gamma(p^2,q^2)$, which we call $\Gamma_{\rm th}(p^2,q^2)$. 
This expression may also be written as a single dispersion representation in $p^2$: 
\begin{eqnarray}
\label{disp_th}
\Gamma_{\rm th}(p^2,q^2)=\int \frac{ds}{s-p^2-i0}\Delta_{\rm th}(s,q^2). 
\end{eqnarray}
Quark--hadron duality \cite{svz,shifman} postulates that $\Gamma_{\rm th}$ and $\Gamma_{\rm phen}$ are equal to each other 
after a proper smearing  is applied to both of them.  
The smearing may be implemented, e.g., by applying the Borel transform \cite{svz}. 
Calculating the Borel image of $\Gamma_{\rm th}$ and $\Gamma_{\rm phen}$ with the parameter 
$2\mu_B^2$, such that $1/(s-p^2)\to \exp(-s/2\mu_B^2)$, we obtain the sum rule for $\Gamma(p^2,q^2)$: 
\begin{eqnarray}
\label{sr_1}
f_{M_Q}\,F_{M_Q\to M}(q^2) \exp\left({-M_Q^2/2\mu_B^2}\right)+
\int {ds}\,\theta(s-s_{\rm cont})\exp\left({-s/2\mu_B^2}\right)\Delta_{\rm phen}(s,q^2)=
\int {ds}\exp\left({-s/2\mu_B^2}\right)\Delta_{\rm th}(s,q^2).\nonumber\\ 
\end{eqnarray}
Let us ignore for a moment the dependence of $\Delta$ on $q^2$ and 
introduce the continuum subtraction point $s_0$ 
(an effective continuum threshold, different from the physical continuum threshold $s_{\rm cont}$) 
according to 
\begin{eqnarray}
\label{s0}
\int  ds\,\theta(s-s_0) \exp\left({-s/2\mu_B^2}\right)\Delta_{\rm th}(s)=
\int  ds\,\theta(s-s_{\rm cont})\exp\left({-s/2\mu_B^2}\right) \Delta_{\rm phen}(s). 
\end{eqnarray}
For sufficiently large $s$, one expects $\Delta_{\rm th}(s)=\Delta_{\rm phen}(s)$ 
(up to oscillating terms \cite{shifman}). 
The region near the continuum threshold $s\simeq s_{\rm cont}$ is 
clearly below the duality region: recall that $\Delta_{\rm phen}(s_{\rm cont})=0$, while 
$\Delta_{\rm th}(\rm s_{cont})>0$. Thus, in the vicinity of $s_{\rm cont}$, 
$\Delta_{\rm th}>\Delta_{\rm phen}$. 
As a result, the continuum subtraction $s_0$ as defined by (\ref{s0}) depends on the Borel parameter 
(and on $q^2$) 
and lies above the physical threshold: $s_0(\mu_B^2)>s_{\rm cont}$.   
Then the sum rule (\ref{sr_1}) takes the form 
\begin{eqnarray}
\label{srff}
f_{M_Q}\,F_{M_Q\to M}(q^2)=
\exp\left({ M_Q^2/2\mu_B^2}\right)\hat \Gamma_{\rm th}(\mu_B^2,q^2,s_0(\mu_B^2,q^2)),   
\end{eqnarray}
with the correlator to which the $s$-cut is applied (hereafter referred to as the cut correlator)
\begin{eqnarray}
\label{3.9}
\hat \Gamma_{\rm th}(\mu_B^2,q^2,s_0)=
\int ds\, \theta(s<s_0)\,\exp\left({-s/2\mu_B^2}\right)\Delta_{\rm th}(s,q^2). 
\end{eqnarray}
If one had known the $\mu_B$- and $q^2$-dependent solution $s_0(\mu_B^2,q^2)$ to (\ref{s0}), 
which encodes the full nonperturbative dynamics, 
the expression (\ref{srff}) would have been exact. Taking $s_0$ constant makes it approximate. 
One may then impose the stability criterium which is expected to guarantee the extraction of the 
true form factor in several different ways (see, e.g., \cite{svz,jamin,tk}). 

We shall now concentrate on different possibilities to calculate $\Gamma_{\rm th}(p^2,q^2)$ and
compare the corresponding results.

\begin{figure}[t]
\begin{center}
\includegraphics[width=12cm]{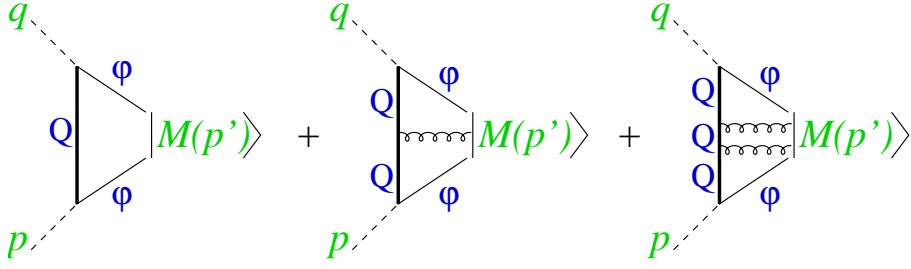}
\end{center}
\caption{\label{Fig:1}Feynman diagrams for the correlator $\Gamma(p,q)$. 
The normal and bold lines denote the heavy $Q$ and the light $\varphi$ particles,
respectively, the wavy line corresponds to the scalar particle which 
mediates the $Q$--$\varphi$ interaction (a scalar analog of the gluon in QCD);   
$|M(p')\rangle$ is the state-vector of the $\varphi\varphi$ bound-state with the momentum $p'$. 
The contribution of the first diagram is expressable through the BS wave function of the light
meson and the full $Q$-propagator. In the region of not very large $q$, $q^2\ll m_Q^2$,  
the second and the third terms are suppressed by higher powers of the heavy-quark mass ($1/m_Q$ and $1/m_Q^2$, respectively) 
compared to the first term and will not be considered.}
\end{figure}

\subsection{Theoretical calculation of $\Gamma(p^2,q^2)$}
We now calculate this correlator using the underlying field-theoretic degrees of freedom, 
and denote the result as $\Gamma_{\rm th}$. The corresponding Feynman diagrams are shown in
Fig.~\ref{Fig:1}. In the region of not too large timelike momentum transfer $q$, $q^2\ll m_Q^2$, the first diagram gives the main contribution whereas 
the contributions of the other diagrams are suppressed by higher powers of the heavy-particle
mass $m_Q$. Neglecting the contributions of the power-suppressed diagrams leads to 
\begin{eqnarray}
\label{3.5a}
\Gamma_{\rm th}(p^2,q^2)=i\int dx \exp({ipx}){\cal D}_{Q}(x)
\langle 0|T \varphi(x)\varphi(0)|M(p')\rangle,  
\end{eqnarray}
where ${\cal D}_Q$ is the full propagator of the heavy quark. 
Approximating the full propagator with the free propagator gives
\begin{eqnarray}
\label{3.5}
\Gamma_{\rm th}(p^2,q^2)=\frac1{(2\pi)^4}\int d^4k d^4x \exp\left({ipx-ikx}\right)\frac{1}{m_Q^2-k^2-i0}
\langle 0|T \varphi(x)\varphi(0)|M(p')\rangle. 
\end{eqnarray}
We can proceed further in two different ways:

A. First, the correlator (\ref{3.5}) may be expressed in terms of the BS amplitude 
in momentum space
\begin{eqnarray}
\Gamma_{\rm th}(p^2,q^2)=\frac1{(2\pi)^4}\int d^4k 
\frac{\Psi_{\rm BS}(k,p')}{m_Q^2-(p-k)^2-i0}. 
\end{eqnarray}
Making use of (\ref{bsmom}) gives
\begin{eqnarray}
\Gamma_{\rm th}(p^2,q^2)=\frac1{(2\pi)^4 i}
\int\limits_0^1 d\xi \int\limits_0^\infty dz ~G(z,\xi)
\int \frac{d^4k}{\left[{z+m^2-\xi(1-\xi)M^2-(k-\xi p')^2-i0}\right]^3
[m_Q^2-(p-k)^2-i0]}. 
\end{eqnarray}
Introducing the Feynman parameter $\alpha$ and performing the $k$-integration, we find
\begin{eqnarray}
\label{3.6}
\Gamma_{\rm th}(p^2,q^2)&=&\frac1{32\pi^2}
\int\limits_0^1 d\xi \int\limits_0^\infty dz ~G(z,\xi)\int\limits_0^1d\alpha (1-\alpha)^2
\frac{1}{\left[\mu^2(1-\alpha)+m_Q^2 \alpha  -P^2\alpha(1-\alpha)\right]^2},\nonumber\\ 
&&\mu^2\equiv z+m^2-\xi(1-\xi)M^2,\qquad P^2=p^2(1-\xi)-\xi(1-\xi)M^2 +q^2\xi.
\end{eqnarray}
From this expression, it is straightforward to obtain the single dispersion representation for 
$\Gamma_{\rm th}$  in the form 
\begin{eqnarray}
\label{disp_th_2}
\Gamma_{\rm th}(p^2,q^2)=\int\limits_{(m_Q+m)^2}^\infty \frac{ds}{(s-p^2-i0)^2}\sigma_{\rm th}(s,q^2),    
\end{eqnarray}
with 
\begin{eqnarray}
\sigma_{\rm th}(s)&=&
\frac1{32\pi^2}
\int\limits_0^1 \frac{d\xi}{(1-\xi)^2} \int\limits_0^\infty dz~ G(z,\xi)
\int\limits_0^1 \frac{d\alpha}{\alpha^2}
\delta(s-s_\alpha), \nonumber\\
&&s_\alpha=\left(\frac{z+m^2-M^2\xi(1-\xi)}{\alpha}+\frac{m_Q^2}{1-\alpha}
\right)\frac{1}{1-\xi}+M^2\xi-q^2\frac{\xi}{1-\xi}.
\end{eqnarray}
The spectral density $\sigma_{\rm th}(s,q^2)$ vanishes at the lower limit of the $s$-integration at
$s=(m_Q+m)^2$. Notice that the quantity $m$ in the lower limit of the $s$-integration
is the mass of the light spectator. For more details about the analytic properties of the
three-point functions, we refer to \cite{lms}.

Performing an integration by parts in (\ref{disp_th_2}), we obtain the spectral representation in the 
``standard'' form (\ref{disp_th}) with 
$\Delta_{\rm th}(s,q^2)=\frac{\partial}{\partial s}\sigma_{\rm th}(s,q^2).$
The Borel images of the two representations (\ref{disp_th_2}) and 
(\ref{disp_th}) read, respectively, 
\begin{eqnarray}
\label{3.17}
\hat \Gamma_{\rm th}(\mu_B^2,q^2)=\frac{1}{2\mu_B^2}
\int\limits_{(m_Q+m)^2}^\infty {ds}\exp(-s/2\mu_B^2)\sigma_{\rm th}(s,q^2)  
\end{eqnarray}
and 
\begin{eqnarray}
\label{3.18}
\hat \Gamma_{\rm th}(\mu_B^2,q^2)=
\int\limits_{(m_Q+m)^2}^\infty {ds}\exp(-s/2\mu_B^2)\Delta_{\rm th}(s,q^2).  
\end{eqnarray}
When the $s$-integration extends to infinity, the equality of both formulas is 
demonstrated by integration by parts. 

We now want to obtain the cut Borel image of Eq.~(\ref{3.9}). 
Recall, that the cut at $s_0$ is applied to the Borel transform of the spectral representation 
for the correlator in the form (\ref{3.5}). 
Then, the relation  
\begin{eqnarray}
\label{surface}
\int\limits_{(m_Q+m)^2}^{s_0}{ds}\exp(-s/2\mu_B^2)\Delta_{\rm th}(s,q^2)=
\frac{1}{2\mu_B^2}\int\limits_{(m_Q+m)^2}^{s_0}{ds}\exp(-s/2\mu_B^2)\sigma_{\rm th}(s,q^2) 
+\exp(-s_0/2\mu_B^2)\sigma_{\rm th}(s_0,q^2)
\end{eqnarray}
leads to the appearance of an additional surface term in the cut spectral representation of the form Eq.~(\ref{3.17}):  
\begin{eqnarray}
\label{gammaborelfullcut}
\label{3.20}
\hat \Gamma_{\rm th}(\mu_B^2,q^2,s_0)&=&
\frac{1}{2\mu_B^2}\int\limits_{(m_Q+m)^2}^{s_0}{ds}\exp(-s/2\mu_B^2)\sigma_{\rm th}(s,q^2) 
+\exp(-s_0/2\mu_B^2)\sigma_{\rm th}(s_0,q^2).
\end{eqnarray} 
Moreover, if one works with the Borel image in the form (\ref{3.20}), the surface term gives the main 
contribution to the correlator for large values of $\mu_B$.

The sum rule for the form factor can now be written in two equivalent ways:  
\begin{eqnarray}
\label{sr_full}
f_{M_Q} F_{M_Q\to M}(q^2)&=&
\frac{1}{2\mu_B^2}\int\limits_{(m_Q+m)^2}^{s_0}{ds}\exp\left(-\frac{s-M_Q^2}{2\mu_B^2}\right)
\sigma_{\rm th}(s,q^2)
+\exp\left(-\frac{s_0-M_Q^2}{2\mu_B^2}\right)\sigma_{\rm th}(s_0,q^2)\\
&=&
\int\limits_{(m_Q+m)^2}^{s_0}{ds}\exp\left(-\frac{s-M_Q^2}{2\mu_B^2}\right)\Delta_{\rm th}(s,q^2) .
\end{eqnarray}

B. Another possibility to calculate $\Gamma_{\rm th}(p^2,q^2)$ is to substitute 
the 
light-cone expansion of $\Psi_{\rm BS}(x,p')$, Eq.~(\ref{lcexpansion}), into Eq.~(\ref{3.5}):
\begin{eqnarray}
\label{3.22}
\Gamma_{\rm th}(p^2,q^2)=
\frac1{(2\pi)^4}\int d^4k\, d^4x \exp\left({ipx-ikx}\right)\frac{1}{m_Q^2-k^2-i0}\sum_{n=0}^{\infty}(x^2)^n\int\limits_0^1
d\xi \exp({-i p'x\xi})\phi_n(\xi,\lambda). 
\end{eqnarray}
The functions $\phi_i(\xi)$ are related to $G(z,\xi)$ via Eqs.~(\ref{2.6}) and (\ref{2.9}).
Performing the $x$ and $k$ integrations gives  
\begin{eqnarray}
\label{3.23}
\Gamma_{\rm th}(p^2,q^2)=
\int\limits_0^1
\frac{d\xi ~\phi_0(\xi,\lambda)}{m_Q^2-(p-\xi p')^2}
-8m_Q^2 \int\limits_0^1\frac{d\xi ~\phi_1(\xi,\lambda)}{\left[m_Q^2-(p-\xi p')^2\right]^3}+\cdots.
\end{eqnarray}
Here $(p-\xi p')^2=p^2(1-\xi)-M^2\xi(1-\xi)+q^2\xi$.
This series (\ref{3.23}) may be written as the series of spectral representations 
\begin{eqnarray}
\label{3.23a}
\Gamma_{\rm th}(p^2,q^2)=
\int\limits_{m_Q^2}^{\infty} 
\frac{ds}{s-p^2}\Delta^{(0)}(s,q^2)+
\int\limits_{m_Q^2}^{\infty}\frac{ds}{(s-p^2)^3}\Delta^{(1)}(s,q^2)+\cdots,
\end{eqnarray}
with $\Delta^{(n)}$ calculable via $\phi_n$. For instance, for $q^2=0$ one finds
\begin{eqnarray}
\label{3.23b}
\Delta^{(0)}(s,q^2=0)=\frac{1}{m_Q^2}(1-\xi^*)\phi_0(\xi^*,\lambda),\quad
\Delta^{(1)}(s,q^2=0)=-8\frac{\phi_1(\xi^*,\lambda)}{1-\xi^*}, \qquad \xi^*=1-m_Q^2/s.
\end{eqnarray}
The corresponding uncut Borel image reads 
\begin{eqnarray}
\label{3.24}
\hat \Gamma_{\rm th}(\mu_B^2,q^2)=
\int\limits_{m_Q^2}^{\infty} 
ds\exp(-s/2\mu_B^2)\Delta^{(0)}(s,q^2)
+
 \frac{1}{2(2\mu_B)^2}\int\limits_{m_Q^2}^{\infty}
ds\exp(-s/2\mu_B^2)\Delta^{(1)}(s,q^2)+\cdots,
\end{eqnarray}
or, equivalently, 
\begin{eqnarray}
\label{3.24a}
\hat \Gamma_{\rm th}(\mu_B^2,q^2)=\int\limits_0^1 \frac{d\xi}{1-\xi}
\left[
\phi_0(\xi)-\frac{m_Q^2}{\mu_B^4}\frac{\phi_1(\xi,\lambda)}{(1-\xi)^2}+\cdots \right]
\exp({-s_\xi/{2\mu_B^2}}), 
\end{eqnarray}
with 
\begin{eqnarray}
s_\xi=\frac{m_Q^2}{1-\xi}+M^2\xi -q^2 \frac{\xi}{1-\xi}.
\end{eqnarray}
In (\ref{3.24}) and (\ref{3.24a}) the dots stand for terms corresponding to higher $n$.  
Clearly, their contributions are suppressed with powers of the Borel parameter $\mu_B$ and vanish in the limit 
$\mu_B\to\infty$. 

Usually, the Borel parameter has the following dependence on the heavy quark mass: $\mu_B^2=m_Q\beta$ \cite{ck}, 
where $\beta$ stays finite in the limit $m_Q\to\infty$. Then, the terms corresponding to higher $n$ are not
suppressed at $m_Q\to\infty$. 

Now, let us consider the cut Borel image of Eq.~(\ref{3.23a}). 
As explained above, the cut is applied to the dispersion representation 
in the form (\ref{3.5}). Therefore, if we want to work with the spectral densities $\Delta^{(i)}$, 
we should take into account the surface terms, similar to Eq.~(\ref{3.20}). 
The cut Borel image takes the form 
\begin{eqnarray}
\label{cutlc}
\hat \Gamma_{\rm th}(\mu_B^2,q^2,s_0)&=&
\int\limits_{m_Q^2}^{s_0} 
{ds}\exp(-s/2\mu_B^2)\Delta^{(0)}(s,q^2)
\\ \nonumber
&&+
\int\limits_{m_Q^2}^{s_0} 
\frac{ds\exp(-s/2\mu_B^2)\Delta^{(1)}(s,q^2)}{2(2\mu_B)^2}
+
\left[
\frac{\partial\Delta^{(1)}(s_0,q^2)}{\partial s_0}
+
\frac{\Delta^{(1)}(s_0,q^2)}{(2\mu_B)^2}\right]\frac{\exp(-s_0/2\mu_B^2)}{2}
\\
\nonumber
&&+\mbox{contributions of terms corresponding to higher $n$}.
\end{eqnarray}
%
Making use of the distribution amplitudes $\phi_n$, we rewrite the correlator (\ref{cutlc}) as 
\begin{eqnarray}
\label{cutlc1}
\hat \Gamma_{\rm th}(\mu_B^2,q^2,s_0)&=&\int\limits_0^1 \frac{d\xi}{1-\xi}
\left[
\phi_0(\xi)-\frac{m_Q^2}{\mu_B^4}\frac{\phi_1(\xi,\lambda)}{(1-\xi)^2}+\cdots \right]
\exp({-s_\xi/{2\mu_B^2}})\theta(s_\xi <s_0)\nonumber\\	
&-&4\exp(-s_0/{2\mu_B^2})
\left[
\frac{\phi_1(\xi_0)}{m_Q^2}
+\frac{\phi'_1(\xi_0)}{4\mu_B^2(1-\xi_0)}+\frac{\phi'_1(\xi_0)}{s_0}
\right]+\cdots, 
\end{eqnarray}
where $\xi_0=1-m_Q^2/s_0$ and $\cdots$ stand for contributions of the terms corresponding 
to $n\ge 2$.

The light-cone correlator corresponds to the first term in the 
expansions (\ref{cutlc}) and (\ref{cutlc1}), calculated with the cut parameter 
$s_0^{\rm LC}$ specific for the LC approximation (see the next section for details): 
\begin{eqnarray}
\label{gammaborellc}
\hat \Gamma_{\rm LC}(\mu_B^2,q^2,s_0^{\rm LC})=
\int\limits_{m_Q^2}^{s_0^{\rm LC}}
{ds}\exp(-s/2\mu_B^2)\Delta^{(0)}(s,q^2)
=\int\limits_0^1 \frac{d\xi}{1-\xi}
\phi_0(\xi)\exp({-s_\xi/{2\mu_B^2}})\theta(s_\xi <s_0^{\rm LC}). 
\end{eqnarray}
Keeping only this term leads to the following sum rule:
\begin{eqnarray}
\label{sr_lc}
f_{M_Q} F_{M_Q\to M}(q^2)= \int\limits_0^1 \frac{d \xi}{1-\xi} 
\phi_0(\xi)\exp\left(-\frac{s_\xi-M_Q^2}{2\mu_B^2}\right)\theta(s_\xi <s_0^{\rm LC} ). 
\end{eqnarray}
Let us study the conditions under which the contributions of higher $n$ are parametrically 
suppressed compared to the contribution of the light-cone $n=0$. 

1. The heavy-meson mass is related to the heavy quark mass by $M_Q=m_Q+\varepsilon_Q$, 
$\varepsilon_Q\simeq \Lambda_{\rm QCD}$. 
One can then check that the dominant contribution to the integral comes from the end-point 
region $\xi\sim \Lambda_{\rm QCD}/m_Q$. Therefore, in order to have contributions of higher $n$ suppressed by powers of 
$1/m_Q$ compared to the $n=0$ contribution, one needs, e.g., the following end-point behavior: 
\begin{eqnarray}
\label{nb}
\phi_n(\xi)\sim \xi^{n+1}. 
\end{eqnarray}
Such a behavior can be obtained, e.g., for the light-cone wave function 
\begin{eqnarray}
\Psi_{\rm LC}(\xi,k_\perp^2)=\exp\left(-{k_\perp^2}/{2\beta_M^2\xi(1-\xi)}\right),
\end{eqnarray}  
$\beta_M$ being the size parameter of the light meson. 
In this specific case the functions $\phi_n(\xi)$ have the necessary behavior (\ref{nb}) in the 
end-point region. 
However, for realistic wave functions obtained 
as solutions to the BS equation, the distribution amplitudes $\phi_n(\xi)$ for different $n$ have a similar behavior 
in the end-point region and therefore the LC expansion for the form factor has no small parameter. 
In this case, the term $n=0$ is not parametrically enhanced compared to terms of higher $n$, and, 
in order to 
obtain the form factor, terms for all $n$ should be summed. 
In other words, the transverse motion of the light 
quark is essential, the longitudinal light-cone distribution amplitude $g_0(\xi)$ is not sufficient, and  
the knowlegde of the full wave function $\Psi(\xi,k_\perp)$ is necessary to obtain the $B\to M$ form factor.

2. One might expect to have a power suppression of the off-LC terms corresponding to $n\ge 1$ 
in the limit $\mu_B\to\infty$, within the ``local-duality'' sum rule \cite{radyushkin}, 
taking into account that in standard SVZ sum rules the choice $\mu_B\to\infty$ suppresses 
condensate contributions. 
Within the context of light-cone sum rules, the limit $\mu_B\to\infty$ is, however, tricky: 
In the uncut light-cone correlator the limit $\mu_B\to\infty$ indeed suppresses the off-LC 
effects ($x^2\ne 0$ terms). However, after applying the cut, this property is lost: obviously,   
the surface terms in the cut correlator (\ref{cutlc}) corresponding to $n\ge 1$ remain finite in 
the limit $\mu_B\to\infty$ and can give a sizeable contribution. 
There is also another subtlety: A specific feature of the local-duality sum rule is that the 
observables are determined 
to great extent by the value of the continuum subtraction point $s_0$. 
Therefore one cannot apply the 
standard sum-rule stability criteria to extract the physical value of an observable. 
Nevertheless, {\it assuming} the duality interval to be process-independent and 
fixing $s_0$ from one observable allows one to predict other
observables.\footnote{We have shown recently that the local-duality sum rule
should provide a good description of the pion elastic form factor for 
all spacelike momentum transfers \cite{lm}.} However, as we shall see in the next section, 
when the standard criterion to fix the continuum subtraction $s_0$ 
is applied, its value for the full and for the LC correlators turn out to be different from each
other.


Finally, we conclude that for the realistic case of the interaction dominated by one-boson 
exchange at short distances, 
the LC contribution does not dominate the cut correlator parametrically, i.e., the off-LC terms 
are not suppressed by any large parameter compared to the LC term. 
To understand how well the LC contribution numerically compares with the full result, 
we calculate the correlator $\hat \Gamma_{\rm th}$, Eq.~(\ref{gammaborelfullcut}), and the 
LC correlator, Eq.~(\ref{gammaborellc}), making use of the BS wave function obtained as 
solution to the BS equation with a realistic potential dominated by a 
one-boson exchange at small separations.


\section{\label{numerical}Numerical results}
In this section we address the case $q^2=0$, and therefore do not write explicitly the argument $q^2$. 
For the BS kernel $G(z,\xi)=m^2 \delta(z)\xi(1-\xi)$ it is straightforward to calculate 
the spectral densities $\Delta_{\rm th}$ and $\Delta_{\rm LC}$. 
We shall analyse the correlators for the light meson in the final state.  
We may therefore set $M=0$ and find 
\begin{eqnarray}
\Delta_{\rm th}(s)&=&\frac{1}{s^3}\left[
(m_Q^2+m^2)\lambda^{1/2}(s,m_Q^2,m^2)+4 m_Q^2 m^2 
\log\left(\frac{s-m_Q^2-m^2+\lambda^{1/2}(s,m_Q^2,m^2)}{2m_Q m}\right)
\right]\theta(s-(m_Q+m)^2),\nonumber\\
\Delta_{\rm LC}(s)&=&\frac{m_Q^2(s-m_Q^2)}{s^3}\theta(s-m_Q^2).  
\end{eqnarray}
Here $\lambda(s,m_Q^2,m^2)=(s-m_Q^2-m^2)^2-4 m_Q^2 m^2$ and $\Delta_{\rm LC}(s)$ is just  
$\Delta^{(0)}(s,q^2=0)$ of Eq.~(\ref{3.23a}). 
Clearly, in the limit $m\to 0$, both quantites coincide. 
Notice the relation 
\begin{eqnarray}
\label{relbor}
\int\limits_{(m_Q+m)^2}^{\infty} ds\,\Delta_{\rm th}(s)=
\int\limits_{m_Q^2}^{\infty} ds\,\Delta_{\rm LC}(s), 
\end{eqnarray}
which can be checked explicitly. 

Let us now discuss the values of the parameters to be used in numerical estimates.

For the heavy-quark mass the situation is obvious: to study the beauty-meson decay, we choose  
the value $m_Q=4.8$ GeV used in light-cone sum-rules \cite{bb,bz}. The same value 
is employed in quark-model calculations \cite{stech}. To describe decay of the charm meson, 
we set $m_Q=1.4$ GeV according to sum rules \cite{bp}. 

The relevant choice of the light-quark mass parameter $m$ requires clarification. 
Our interest is to understand with the help of the
simplified model under consideration the corresponding QCD calculation. We therefore 
choose the numerical parameters relevant for QCD. 
Since the light-quark mass parameter $m$ appears in the framework of the BS equation, it should be
understood as the effective quark mass which takes into account nonperturbative effects related to
spontaneous chiral symmetry breaking in the soft region, i.e., the constituent quark mass. 
In \cite{ms,msl}, the constituent quark mass was calculated through the
quark condensate in QCD with the result $m\simeq 220$ MeV at the chiral-symmetry breaking scale 
$1$ GeV. The scale-dependence of the constituent mass of the light quark above 1 GeV was also 
reported in \cite{ms}. In sum-rule analyses \cite{bb,bz} it was found, that the relevant infrared
factorization scale for a heavy-meson decay is $\sqrt{M_Q^2-m_Q^2}$. We therefore make use of the
constituent quark mass evaluated at this scale. Employing the results from \cite{ms}, we find 
the relevant value of the constituent mass of the light quark $m=150$ MeV for beauty-meson decay, and 
$m=200$ MeV for charm-meson decay. These values of the quark masses will be used in numerical estimates
below. 

\begin{figure}[b]
\begin{center}
\begin{tabular}{lr}
\includegraphics[width=8cm]{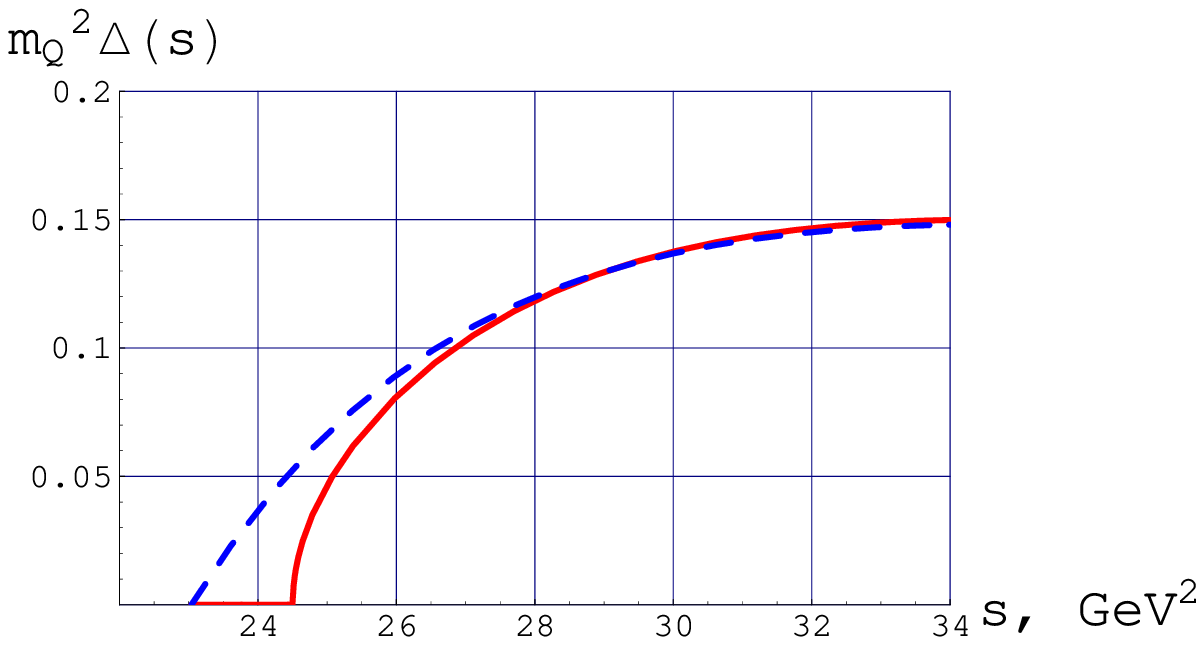}
&
\includegraphics[width=8cm]{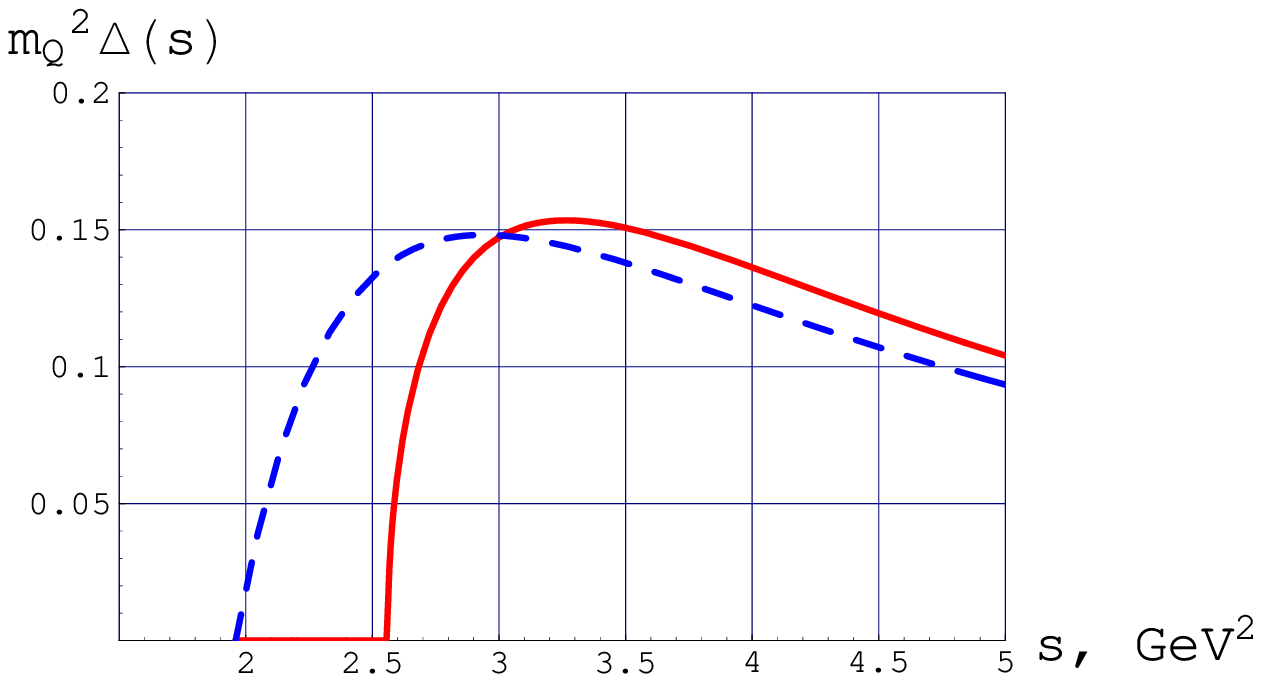} 
\end{tabular} 
\caption{\label{Fig:2}Spectral densities $m_Q^2\Delta_{\rm th}(s)$ (solid red line) and $m_Q^2\Delta_{\rm LC}(s)$ (dashed blue line)
for parameter values corresponding to beauty-meson decay, 
$m_Q=4.8$ GeV, $m=150$ MeV (left), and to charm-meson decay, $m_Q=1.4$ GeV, $m=200$ MeV (right).} 
\end{center}
\end{figure}

Fig.~\ref{Fig:2} shows the quantities $m_Q^2\Delta_{\rm th}$ and $m_Q^2\Delta_{\rm LC}$ 
for two cases: (a) $m_Q=4.8$ GeV and $m=150$ MeV relevant for $B$-decay  and (b) 
$m_Q=1.4$ GeV and $m=200$ MeV relevant for $D$-decay.  
Important for the following is that the thresholds in $\Delta_{\rm th}$ and $\Delta_{\rm LC}$ 
do not coincide: in the light-cone spectral density the threshold is $m_Q^2$ whereas in the
full spectral density it is $(m_Q+m)^2$. The region near the threshold 
provides the main contribution to the cut Borel-transformed correlators, therefore
the mismatch of the thresholds is responsible for the nonvanishing of the off-light-cone 
effects in sum rules. 

Let us briefly address the {\it uncut} Borel-transformed full and LC correlators  
\begin{eqnarray}
\hat \Gamma_{\rm th}(\mu_B^2)=\int\limits_{(m_Q+m)^2}^{\infty} ds\,\exp(-s/2 \mu_B^2)\,\Delta_{\rm th}(s),\qquad
\hat \Gamma_{\rm LC}(\mu_B^2)=\int\limits_{m_Q^2}^{\infty} ds\,\exp(-s/2 \mu_B^2)\Delta_{\rm LC}(s).  
\end{eqnarray}
Eq.~(\ref{relbor}) shows that for large values of the Borel mass $\mu_B$ these two quantites coincide. 
The same conclusion may be obtained directly from Eq.~(\ref{3.24}): 
the terms proportional to the off-LC distribution amplitudes are suppressed by powers of the Borel 
parameter $\mu_B^2$. Thus, in the case of the uncut Borel transform there is a clear limit --- large 
values of the Borel parameter --- in which the off-LC effects vanish, independently of the 
specific value of the light-quark mass. However, the uncut Borel-transformed correlators contain
contributions of all possible hadronic states containing the heavy quark, and therefore 
cannot provide information on the properties of the single heavy meson.  

\begin{figure}[t]
\begin{center}
\begin{tabular}{ll}
\includegraphics[width=8.5cm]{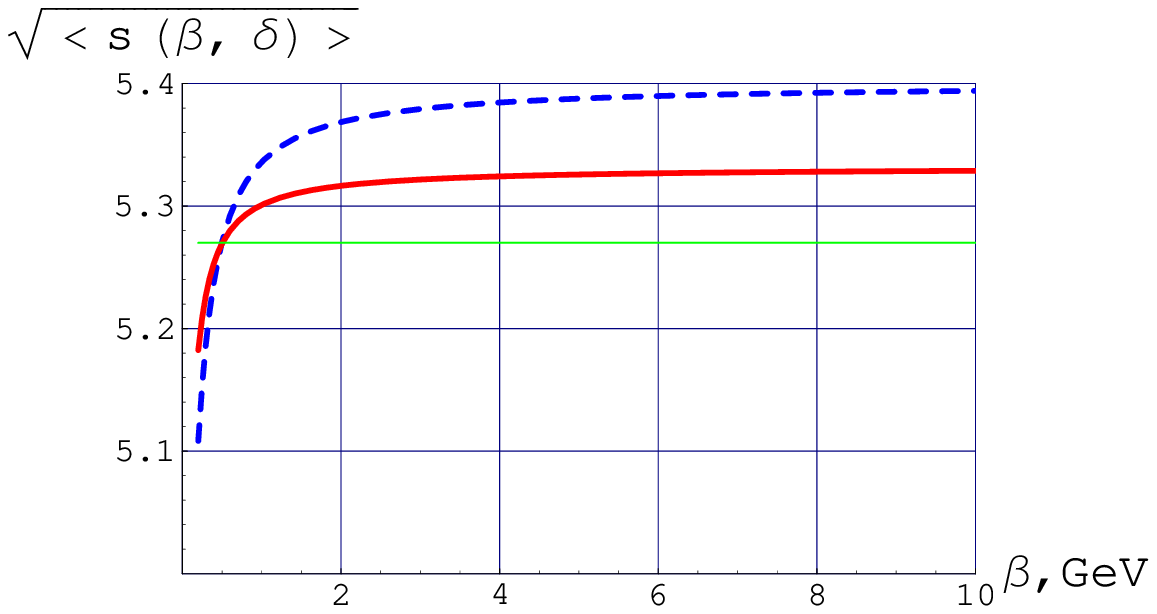}&\includegraphics[width=8.5cm]{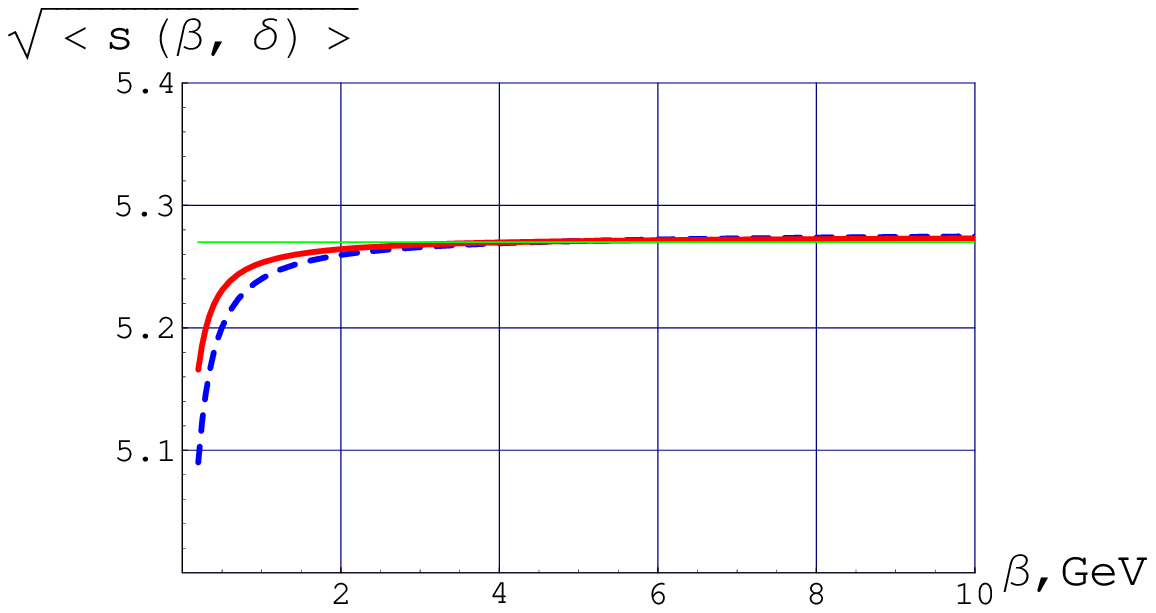}\\
\includegraphics[width=8.5cm]{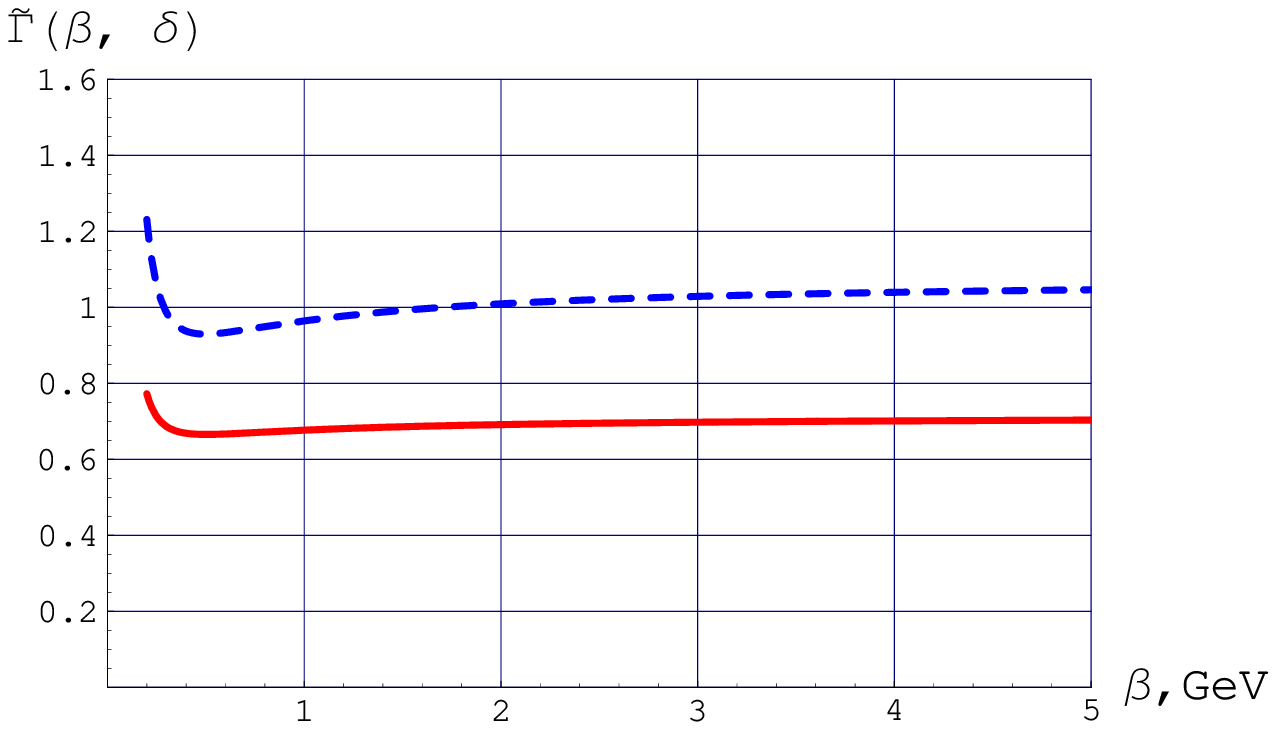}&\includegraphics[width=8.5cm]{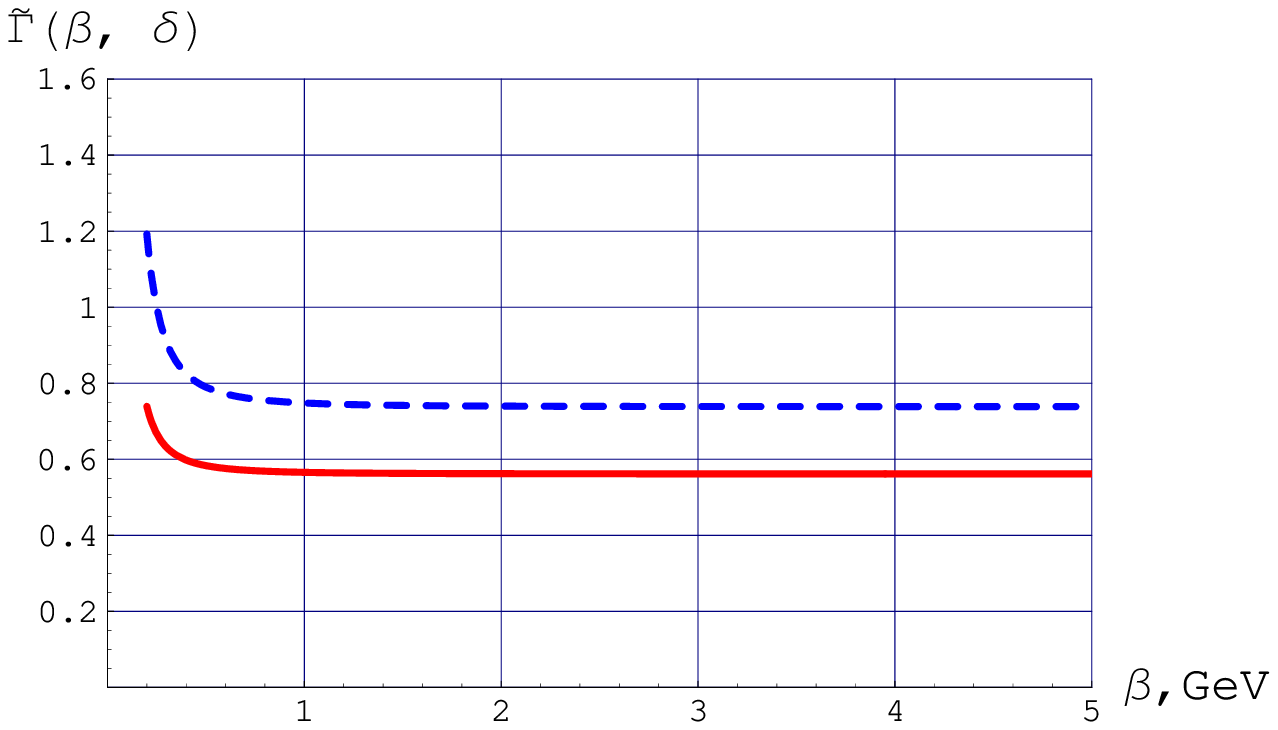}\\
\hspace{.3cm}\includegraphics[width=8.5cm]{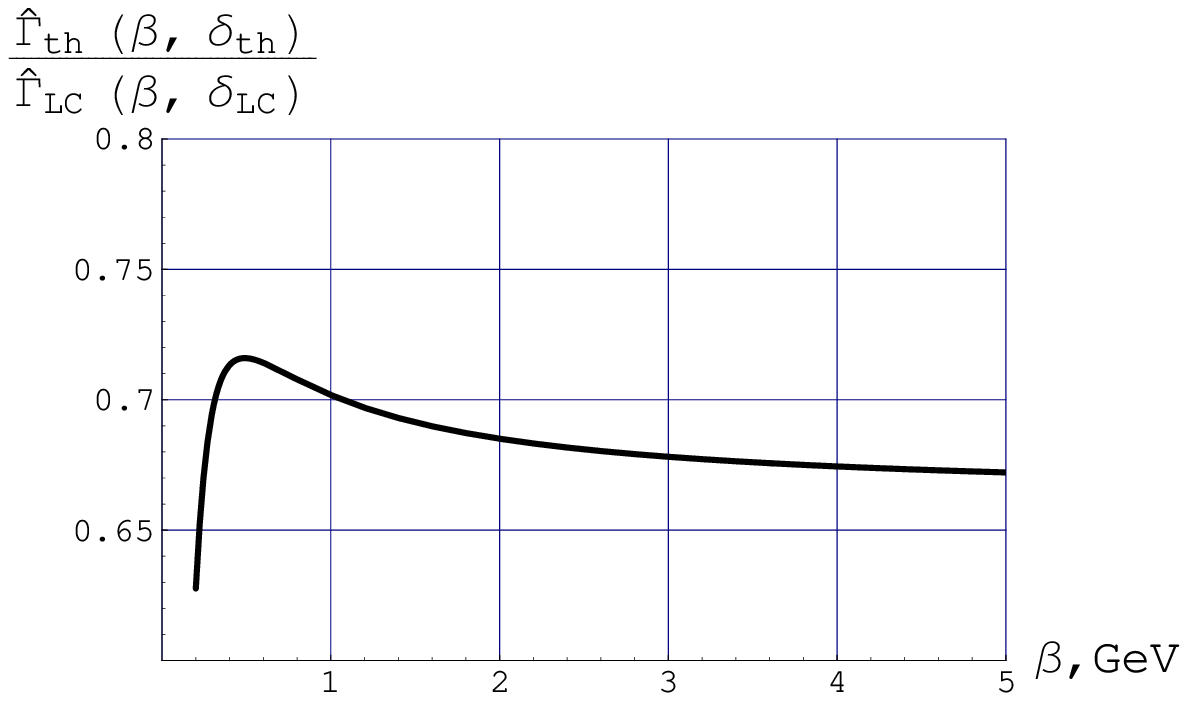}&\hspace{.3cm}\includegraphics[width=8.5cm]{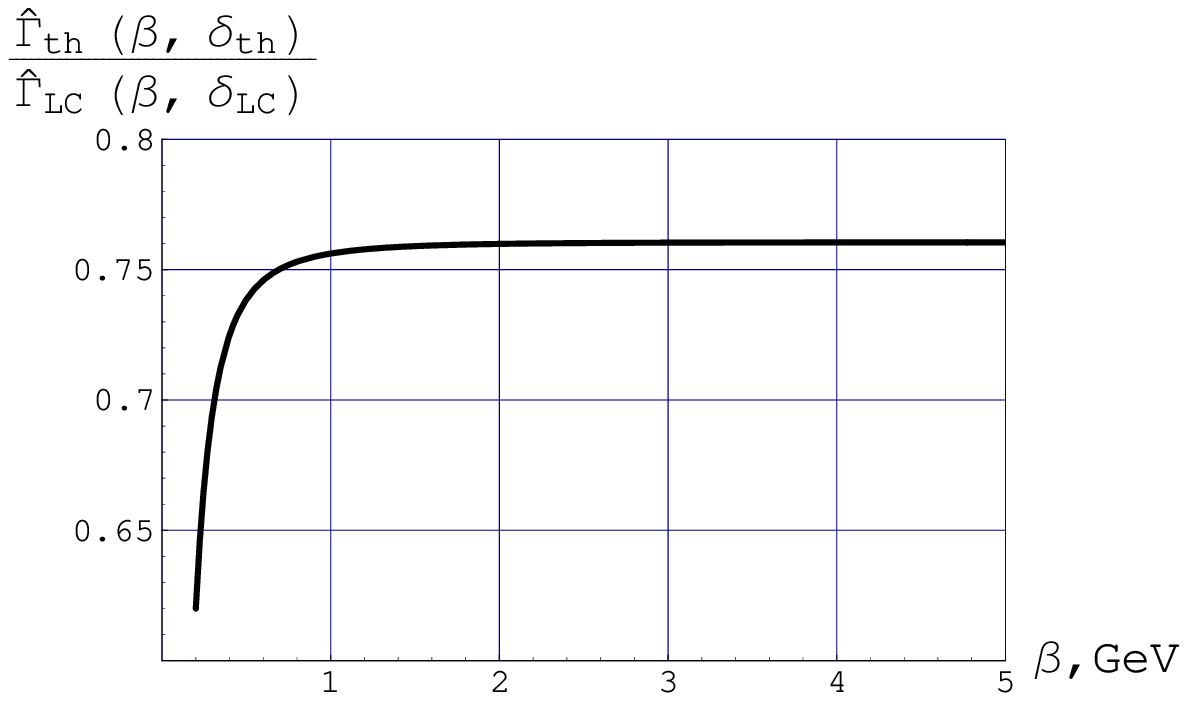}
\end{tabular} 
\caption{\label{Fig:3}Plots for the parameters corresponding to beauty-meson decay 
$m_Q=4.8$ GeV, $m=150$ MeV and $\delta$ fixed by tuning 
$\sqrt{\langle s \rangle}$ to $M_Q=5.27$ GeV at two different values of $\beta$: 
The plots in the left column correspond to $\delta_{\rm LC}=0.96$ GeV and $\delta_{\rm th}=0.79$ 
GeV, which
are obtained from the relation $\sqrt{\langle s \rangle_{\rm LC}}=\sqrt{\langle s \rangle_{\rm th}}=5.27$ GeV 
for $q^2=0$ and $\beta=0.5$ GeV;   
the plots in the right column correspond to 
$\delta_{\rm LC}=0.755$ GeV and $\delta_{\rm th}=0.69$ GeV,  
which are obtained from the relation $\sqrt{\langle s \rangle_{\rm LC}}=\sqrt{\langle s \rangle_{\rm th}}=5.27$ GeV 
for $q^2=0$ and $\beta=4$ GeV.  
{\bf First row}: $\sqrt{\langle s \rangle_{\rm th}}$ (solid red line)
and $\sqrt{\langle s \rangle_{\rm LC}}$ vs $\beta$ (dashed blue
line). The horizontal (green) line is $M_Q=5.27$ GeV.  
{\bf Second row}: $\widetilde\Gamma(\beta, q^2,\delta)$ vs $\beta$ [(\ref{gammas})] at $q^2=0$:  
$\widetilde\Gamma_{\rm th}(\beta, q^2,\delta_{\rm th})$ (solid red line)
and 
$\widetilde\Gamma_{\rm LC}(\beta, q^2,\delta_{\rm th})$ (dashed blue line). 
{\bf Third row:} The ratio 
$\hat\Gamma_{\rm th}(\beta, q^2,\delta_{\rm th})/\hat\Gamma_{\rm LC}(\beta, q^2,\delta_{\rm LC})$
vs $\beta$ for $q^2=0$.}
\end{center}
\end{figure}
Now, let us study the {\it cut} Borel-transformed correlators which are relevant for the 
extraction of the form factors within QCD sum rules. We shall see that for the cut Borel-transformed correlator the 
situation is different: namely, there is no physical limit in which the
off-LC effects are negligible.\footnote{As we have seen, the full and the LC spectral densities 
coincide in the limit $m\to 0$. However, the parameter $m$ should be identified 
with the effective quark mass, which stays finite of order $\Lambda_{\rm QCD}$ in the chiral limit. 
Therefore,  
the limit $m\to 0$ does not correspond to a realistic situation.}
\begin{figure}[t]
\begin{center}
\begin{tabular}{ll}
\includegraphics[width=8.4cm]{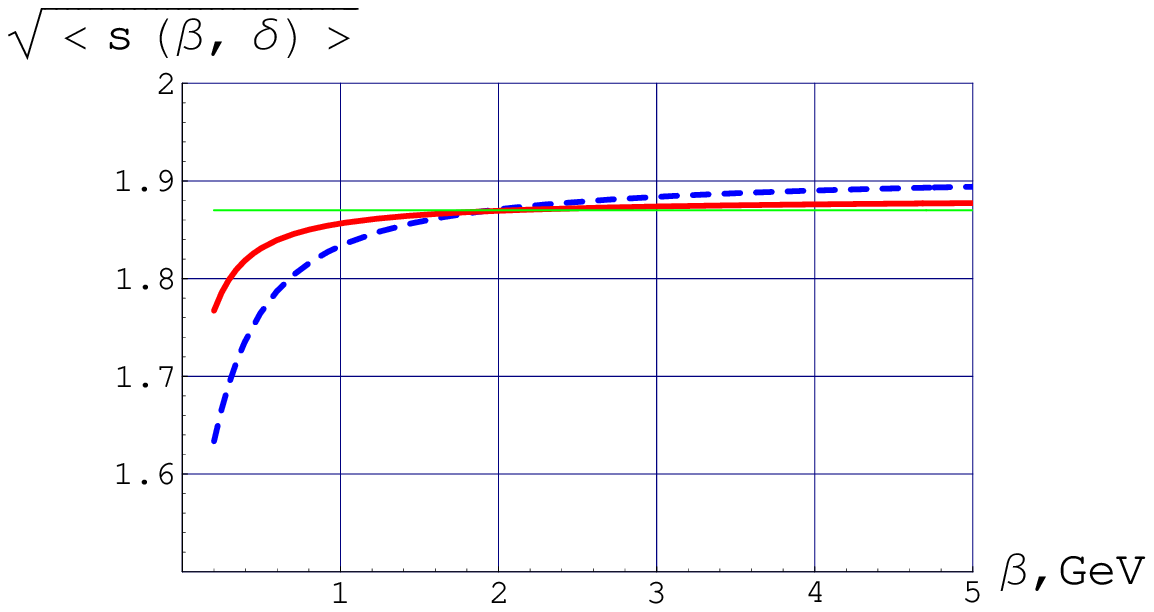}&\includegraphics[width=8.4cm]{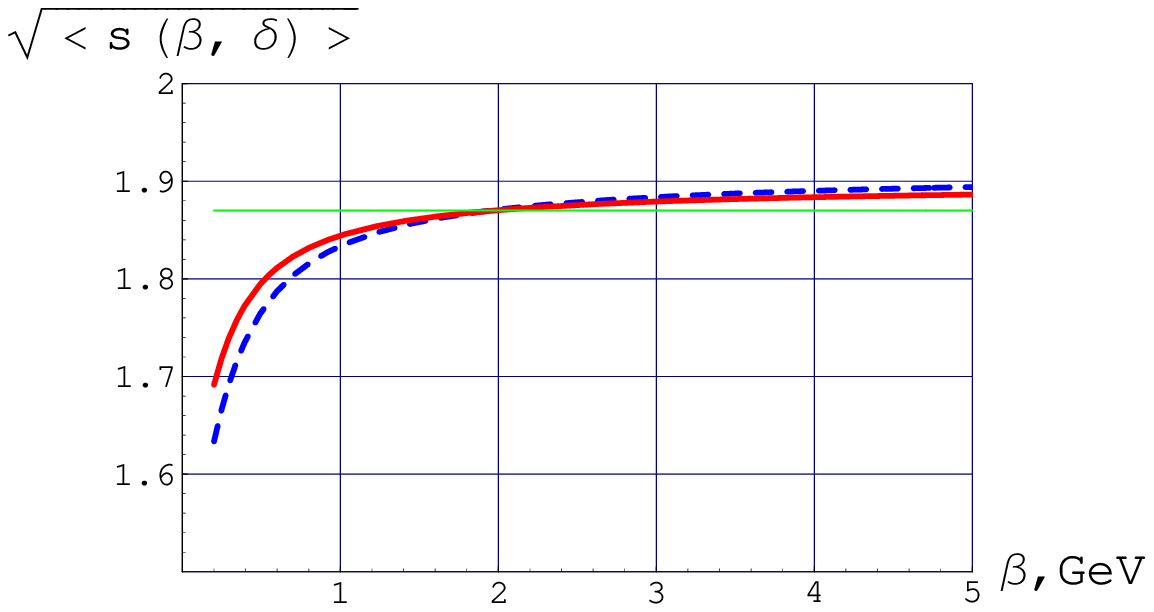}\\
\includegraphics[width=8.4cm]{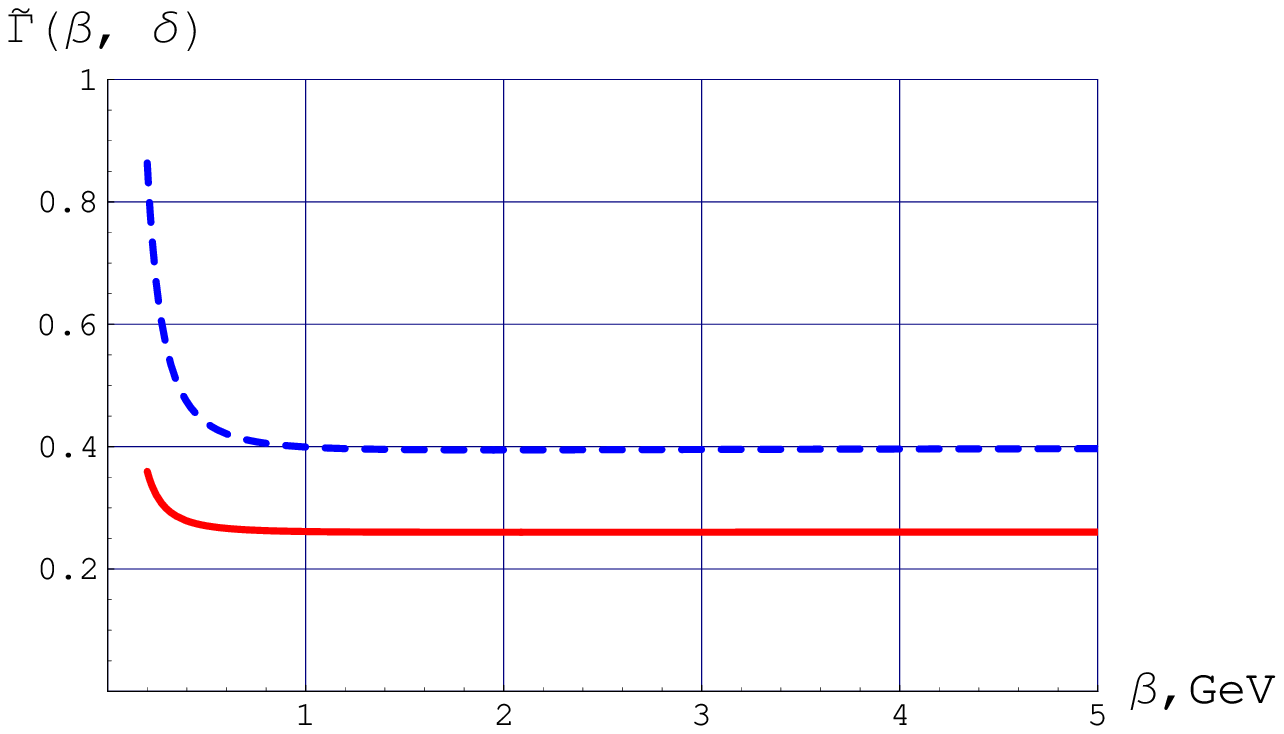}&\includegraphics[width=8.4cm]{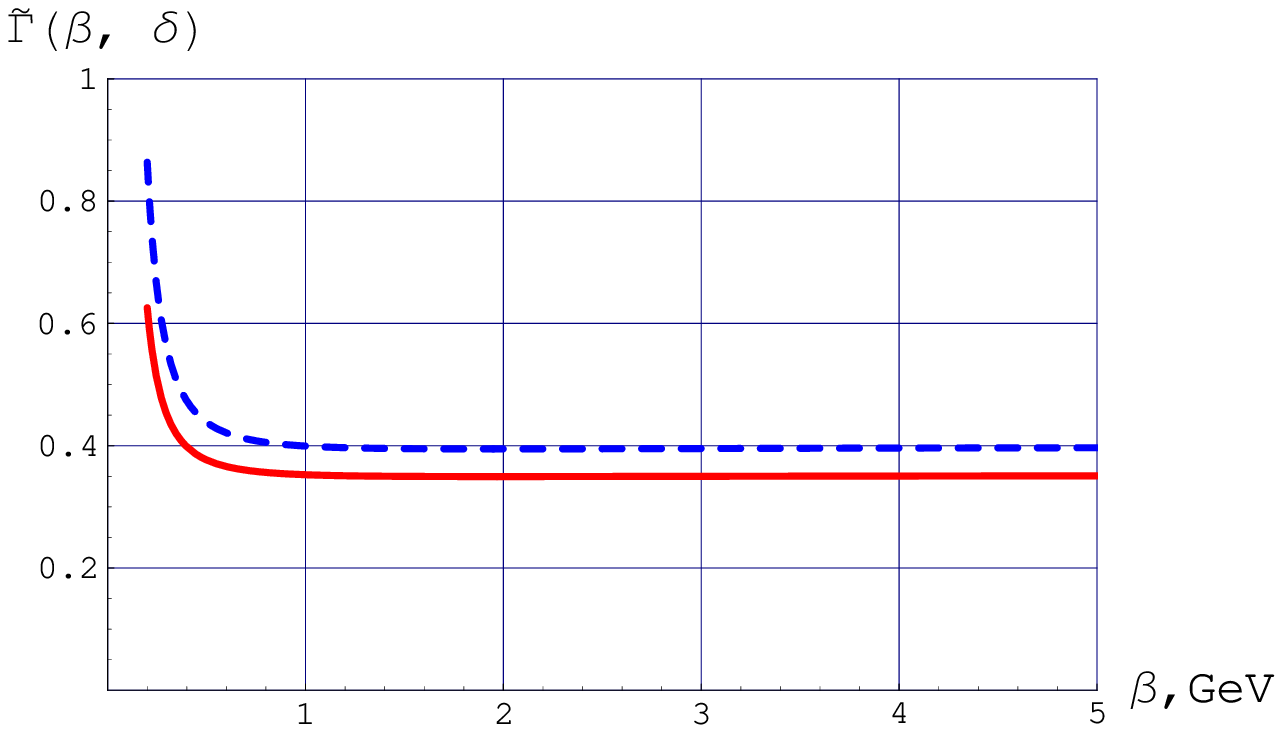}\\
\hspace{.3cm}\includegraphics[width=8.4cm]{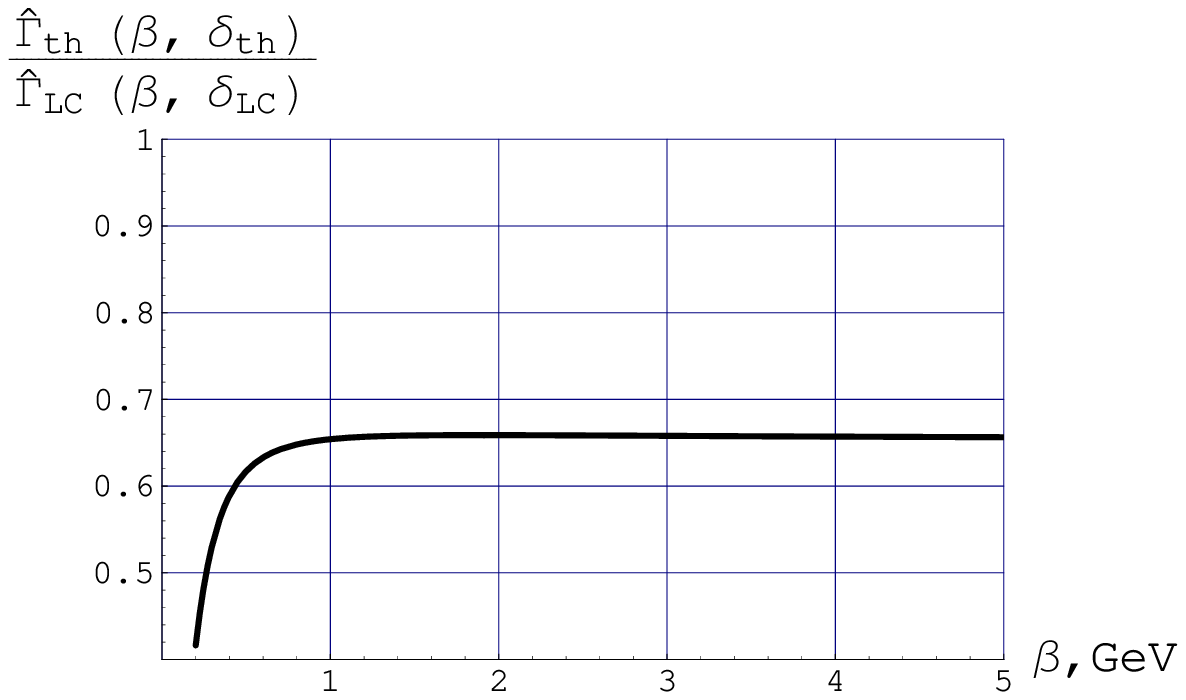}&\hspace{.3cm}\includegraphics[width=8.4cm]{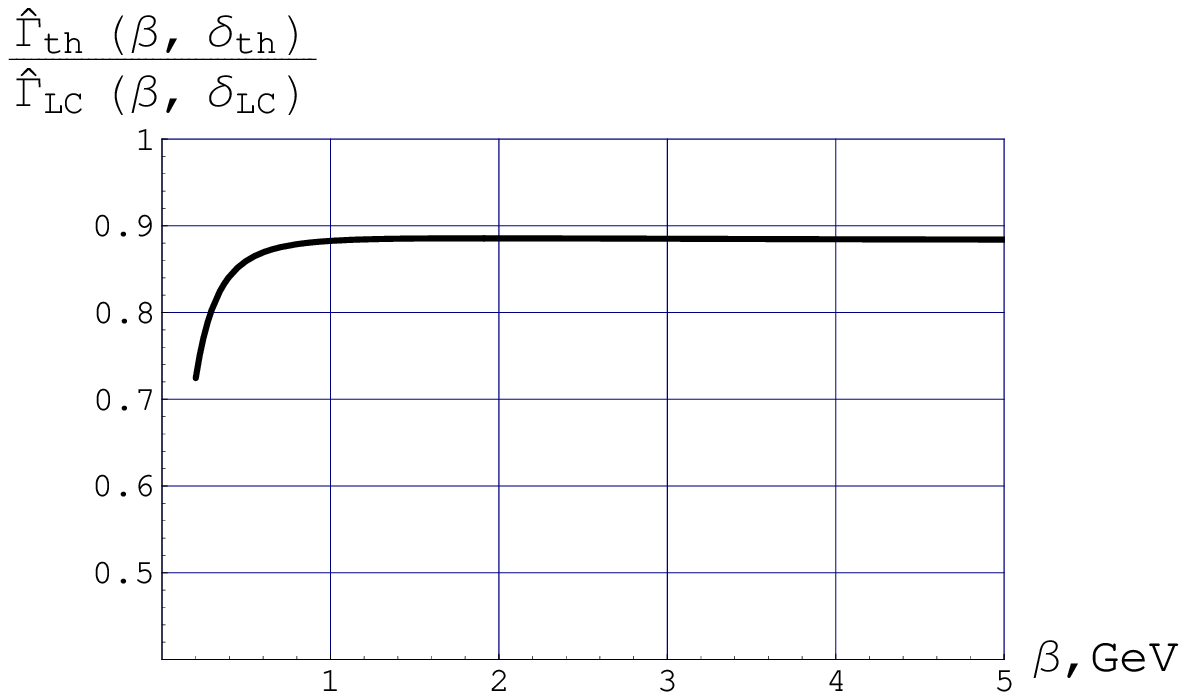}
\end{tabular} 
\caption{\label{Fig:4}Plots corresponding to charm-meson decay $m_Q=1.4$ GeV, 
for $\delta$ fixed by tuning $\sqrt{\langle s\rangle}$ to $M_Q=1.87$ GeV at $\beta=2$ GeV, 
and for two different values of the light-quark mass: 
$m=200$ MeV (left column) and $m=100$ MeV (right column). 
{\bf First row:} $\sqrt{\langle s \rangle_{\rm LC}}$ and $\sqrt{\langle s \rangle_{\rm th}}$ [Eq.~(\ref{delta})] 
vs $\beta$. The horizontal (green) line is $M_Q=1.87$ GeV. 
The parameters $\delta_{\rm LC}=0.93$ GeV and 
$\delta_{\rm th}=0.72$ GeV for $m=200$ MeV (left column) and 
$\delta_{\rm th}=0.85$ GeV for $m=100$ MeV (right column) 
are obtained by requiring that at $q^2=0$ and $\beta=2$ GeV 
$\sqrt{\langle s \rangle_{\rm LC}}=\sqrt{\langle s \rangle_{\rm th}}=1.87$ GeV. 
{\bf Second row:} 
$\widetilde\Gamma(\beta, q^2,\delta)$ [Eq.~(\ref{gammas})] at $q^2=0$: 
$\widetilde\Gamma_{\rm th}(\beta, q^2,\delta_{\rm th})$ (solid red line)
and 
$\widetilde\Gamma_{\rm LC}(\beta, q^2,\delta_{\rm LC})$ (dashed blue line). 
{\bf Third row:} The ratio 
$\hat\Gamma_{\rm th}(\beta, q^2,\delta_{\rm th})/\hat\Gamma_{\rm LC}(\beta, q^2,\delta_{\rm LC})$ at 
$q^2=0$.}
\end{center}
\end{figure}
We introduce the parameters $\delta$ related to the continuum
subtraction $s_0$ by 
\begin{eqnarray}
s_0=(m_Q+\delta)^2. 
\end{eqnarray}
To fix $\delta$, we follow the standard procedure \cite{bz}: namely, we require that the quantity 
\begin{eqnarray}
\label{delta}
\langle s(\beta,q^2,\delta) \rangle \equiv
\frac{\displaystyle\int\limits_{s_{\rm low}}^{(m_Q+\delta)^2}ds\exp\left(-\frac{s-M_Q^2}{2\mu_B^2}\right) \,s\, \Delta(s,q^2)}
{\displaystyle\int\limits_{s_{\rm low}}^{(m_Q+\delta)^2}ds \exp\left(-\frac{s-M_Q^2}{2\mu_B^2}\right)\,\Delta(s,q^2)}\,,
\qquad \mu_B^2=m_Q\beta, 
\end{eqnarray}
reproduces the heavy-meson mass, i.e., 
\begin{eqnarray}
\label{delta2}
\langle s(\beta,q^2,\delta) \rangle=M_Q^2, 
\end{eqnarray}
both for the LC and the full spectral densities, 
where $s_{\rm low}=m_Q^2$ for the LC correlator and $s_{\rm low}=(m_Q+m)^2$ 
for the full correlator. 
The quantity $\delta$ as defined by (\ref{delta2}) depends on $q^2$ and $\mu_B$: 
Eq.~(\ref{delta2}) is just the definition of the implicit function
$\delta(q^2,\mu_B^2)$. Such a procedure of fixing $\delta$ for the light-cone correlator 
was employed e.g. in \cite{bz}. 
Recall, however, that there is no unique way to fix $\delta$: one may 
require instead that $\langle s^n \rangle=(M_Q^2)^n$ for $n>1$. 
Moreover, since the spectral densities and the thresholds in Eq.~(\ref{delta}) are different
for $\delta_{\rm th}$ and $\delta_{\rm LC}$, also the numerical values of  
$\delta_{\rm th}$ and $\delta_{\rm LC}$ obtained from Eq.~(\ref{delta2}) are different. 
We discuss here only the case $q^2=0$. Taking into account the lack of a unique way to 
introduce $\delta$, we shall not consider the $\mu_B$-dependent $\delta_{\rm th}$ and 
$\delta_{\rm LC}$. We rather determine the constant values $\delta_{\rm th}$ and $\delta_{\rm LC}$ 
such that the relation (\ref{delta2}) is satisfied only for one specific value of $\beta$. 

Fig.~\ref{Fig:3} presents the numerical results for beauty-meson decay: $M_Q=5.27$ GeV$, 
m_Q=4.8$ GeV and $m=150$ MeV. We plot the Borel curves for the full and for the LC correlators for 
two different values of $\delta$:   
The left column shows the results for $\delta_{\rm th}=0.86$ GeV and $\delta_{\rm LC}=0.96$ GeV. 
In this case, the relation (\ref{delta2}) is fulfilled at the relatively low value $\beta=0.5$ GeV: 
namely, 
$\sqrt{\langle s \rangle_{\rm LC}}=\sqrt{\langle s \rangle_{\rm th}}=5.27$ GeV 
for $q^2=0$ and $\beta=0.5$ GeV.  
The right column present the results for $\delta_{\rm th}=0.72$ GeV and $\delta_{\rm LC}=0.755$ GeV. 
In this case,  
$\sqrt{\langle s \rangle_{\rm LC}}=\sqrt{\langle s \rangle_{\rm th}}=5.27$ GeV 
for $q^2=0$ and $\beta=4$ GeV. 
The first row shows $\sqrt{\langle s(\beta,\delta)\rangle}$ calculated with the LC and the 
full correlators vs $\beta$. 
The second row presents the quantity 
\begin{eqnarray}
\label{gammas}
\widetilde \Gamma(\beta,q^2,\delta)=m_Q^2\exp\left(\frac{M_Q^2}{2\mu_B^2}\right)
\hat \Gamma(\mu_B^2,q^2,s_0), \quad \mu_B^2=m_Q\beta, \quad s_0=(m_Q+\delta)^2,  
\end{eqnarray}
for the LC and the full correlators. 
Finally, the third row gives the ratio 
of the full to the LC correlators.

Fig.~\ref{Fig:4} gives the results for charm-meson decay: $M_Q=1.87$ GeV, $m_Q=1.4$ GeV and 
$m=200$ MeV (left column). 
To illustrate the influence of the light-quark mass on the off-LC effects, we 
present also the results for $m=100$ MeV (right column). 
The continuum subtraction parameter $\delta$ is fixed from the relation 
$\sqrt{\langle s(\beta,\delta)\rangle}=M_Q$ at $\beta=2$ GeV. 

To show the origin of the difference between the cut full and light-cone correlators, 
we consider the limit $m_Q\to \infty$ and $\mu_B\to\infty$, with
$\mu_B\gg m_Q$.  
In this case explicit expressions for the correlators may be obtained:
\begin{eqnarray}
\label{4.8}
m_Q^2\hat \Gamma_{\rm LC}(\mu^2_B\to\infty,q^2=0,\delta_{\rm LC})&=&
2\delta^2_{\rm LC}+O(\delta^3_{\rm LC}/m_Q),\nonumber \\
m_Q^2\hat \Gamma_{\rm th}(\mu^2_B\to\infty,q^2=0,\delta_{\rm th})&=&
2\delta^2_{\rm th}-m^2 \left[\log\left(\frac{4 \delta^2_{\rm th}}{m^2}\right)+1\right]+O(m^4/\delta^2_{\rm th})
+O(\delta^3_{\rm th}/m_Q).
\end{eqnarray}
For $\mu_B\gg m_Q$ ($\beta\sim m_Q$), the uncut and the cut correlators (both full and LC) 
behave quite differently for large $m_Q$: 
\begin{eqnarray}
\Gamma(\mu_B^2\to\infty,m_Q^2)=O(1), \qquad 
\Gamma(\mu_B^2\to\infty,m_Q^2,\delta)=O(\delta^2/m_Q^2).  
\end{eqnarray}
Thus, the cut correlator picks up only a small
fraction of the full correlator from the region not far from the threshold. 
(For $\beta \ll m_Q$, both the
cut and the uncut correlators have a similar behavior $\sim 1/m_Q^2$.)

Now, fixing 
$\delta_{\rm th}$ and $\delta_{\rm LC}$ according to the standard procedure (\ref{delta2}), 
we express these quantities via the binding energy of the heavy meson $\varepsilon_Q$ defined
according to 
$
M_Q=m_Q+\varepsilon_Q$:\footnote{
Notice that $\delta_{\rm LC}$ turns out to be different 
from $\delta_{\rm th}$. This has the following origin: 
If we include $N$ terms in the LC expansion of the cut correlator (\ref{cutlc}), cut them at 
$\delta^{(N)}_{\rm LC}$, and determine the latter from Eqs.~(\ref{delta}) and (\ref{delta2}),
then 
$\lim\limits_{N\to\infty}\delta^{(N)}_{\rm LC}=\delta_{\rm th}$. 
Since we have included only one $(n=0)$ term, we obtain $\delta^{(0)}_{\rm LC}\ne \delta_{\rm th}$.}
\begin{eqnarray}
\label{deltas}
\delta_{\rm LC}&=&\frac32 \varepsilon_Q,\nonumber\\
\delta_{\rm th}&=&\frac32 \varepsilon_Q-\frac{2m^2}{3\varepsilon_Q}\left[
\log\left(\frac{3 \varepsilon_Q}{m}\right)-1
\right]+\cdots, 
\end{eqnarray}
\vspace{-.5cm}
leading to 
\begin{eqnarray}
m_Q^2\hat \Gamma_{\rm LC}(\mu^2_B\to\infty,q^2=0,\delta_{\rm LC})&=&
\frac92\varepsilon_Q^2,\nonumber \\
m_Q^2\hat \Gamma_{\rm th}(\mu^2_B\to\infty,q^2=0,\delta_{\rm th})&=&
\frac92\varepsilon_Q^2-6m^2\log\left(\frac{3\varepsilon_Q}{\sqrt{e}m}\right)+\cdots, 
\end{eqnarray}
\vspace{-.4cm}
and thus 
\begin{eqnarray}
\frac{\hat \Gamma_{\rm th}(\mu^2_B\to\infty,q^2=0,\delta_{\rm th})}
{\hat \Gamma_{\rm LC}(\mu^2_B\to\infty,q^2=0,\delta_{\rm LC})}=
1-\frac{4m^2}{3\varepsilon_Q^2 }\log\left(\frac{3\varepsilon_Q}{\sqrt{e}m}\right)+\cdots.  
\end{eqnarray}
In the expressions above the dots denote terms containing higher powers of $m/\varepsilon_Q$. 
This example illustrates that the off-LC effects may play an essential role in the cut correlators, as their contribution is
not suppressed by any large parameter: the quantites $m$ and $\varepsilon_Q$ have the same order of magnitude. 

Let us emphasize that we compare the full and the light-cone correlators 
evaluated at different values of the cut parameters $\delta_{\rm LC}$ and $\delta_{\rm th}$. 
From our point of view this very comparison is relevant if one wants to understand 
the error due to taking into account only the light-cone $(x^2=0)$ contribution 
to the correlator and neglecting terms containing higher powers of $x^2$. 

One could also compare the correlators for the same value of $\delta$. The difference between 
the full and the LC correlators is only slightly reduced in this case, the ratio 
still remaining $O(m^2/\delta^2)$. This can be seen by 
setting $\delta_{\rm LC}=\delta_{\rm th}$ in (\ref{4.8}). 

The following lessons may be drawn from the results presented in this section:
\begin{itemize}
\item[a.]
The off-LC effects play an essential role in the cut correlator, as they are not suppressed by any large
parameter. Numerically, the difference between the full and the LC correlators,  
evaluated at the same value of the Borel parameter, is 10$\div$20\%.   
This difference is due to the off-LC effects. 
\item[b.] The functions 
$\widetilde\Gamma_{\rm LC}(\beta)$ and $\widetilde\Gamma_{\rm th}(\beta)$ 
have the same shape, but $\widetilde\Gamma_{\rm LC}(\beta)$ lies well above 
$\widetilde\Gamma_{\rm th}(\beta)$, if the standard
procedure of fixing $\delta_{\rm LC}$ and $\delta_{\rm th}$ (\ref{delta2}) is used. 
The values of both (th and LC) correlators obtained with $\delta_{\rm th}$ and $\delta_{\rm LC}$ 
tuned at $\beta=0.5$ GeV (left column in Fig.~\ref{Fig:3}) 
are greater than the values of the respective correlators 
obtained with $\delta_{\rm th}$ and $\delta_{\rm LC}$ tuned at $\beta=4$ GeV (right column in Fig.~\ref{Fig:3}). 
The local stability is better when one fixes $\delta$ from (\ref{delta2}) at a larger value 
of $\beta$ (right column in Fig. \ref{Fig:3}). 
In this case, both Borel curves for 
$\widetilde\Gamma_{\rm LC}$ and $\widetilde\Gamma_{\rm th}$ show a 
good stability in $\beta$. Nevertheless, still  
$\widetilde\Gamma_{\rm LC}$ is much larger than $\widetilde\Gamma_{\rm th}$! This illustrates that the Borel 
stability {\it per se} does not guarantee the extraction of the correct physical value. 

\item[c.]
The difference between $\widetilde\Gamma_{\rm LC}$ and $\widetilde\Gamma_{\rm th}$
increases with increasing mass of the light quark. 
Therefore, this difference is expected to be greater for the heavy mesons $B_s$ and $D_s$, 
containing the strange $s$-quark, than for $B$ and $D$. 
\end{itemize}


\section{\label{conclusions}Conclusions}
In this paper, we studied the correlator 
$$
i\int dx \exp(ipx)\langle 0|T \varphi(x)Q(x) Q(0)\varphi(0)|M(p')\rangle, 
$$ 
which is one of the basic objects for extracting the heavy-to-light form factor within the 
method of sum rules. 
We have shown that to leading $1/m_Q$-accuracy this correlator may be calculated through 
the BS amplitude of the light meson
\vspace{-.3cm}
$$
\langle 0|T \varphi(x)\varphi(0)|M(p')\rangle.  
$$
Expanding the BS amplitude near the light cone $x^2=0$ 
generates the light-cone expansion of the correlator. 

Making use of the Nakanishi representation for the BS amplitude,\footnote{The Nakanishi
representation leads to technical simplifications, but conceptually
any other form of the BS amplitude may be used.} 
we obtained 
dispersion representations for the full and the 
LC correlators in terms of the kernel $G(z,\xi)$ of the Nakanishi representation. 
We studied the full and the light-cone correlators and their Borel transforms 
depending on the properties of the kernel $G(z,\xi)$.

We then made use of the known solution for $G(z,\xi)$ in a model with 
light scalar particles interacting by an exchange of a massless boson. This relatively simple 
model provides a good laboratory for studying QCD since the corresponding bound-state wave 
functions have properties similar to the properties of hadron wave functions in QCD. 
We calculated the full and the light-cone correlators and their 
Borel transforms in the variable $p^2$, and studied these correlators for various  
prescriptions to fix the heavy-hadron continuum subtraction and in various regions 
of the parameters relevant for extracting the heavy-to-light form factors. 
This work thus represents the 
first systematic study of the off-light-cone effects in QCD sum rules for heavy-to-light form
factors. 

Our main results may be summarized as follows: 
\begin{itemize}
\item[1.]
We have seen that --- after performing the Borel transform --- the light-cone 
correlator provides numerically the bulk of the full 
correlator, although parametrically the off-LC effects are not suppressed 
compared to the LC contribution. This observation holds for various prescriptions of fixing the heavy-hadron continuum subtraction point 
(i.e., a cut applied to the correlator for isolating the contribution of the heavy hadron of interest in the initial state) and a wide range of masses of particles involved in the decay process.  
\item[2.]
We demonstrated that, nevertheless, the difference between the cut full and the cut 
light-cone correlators always remains nonvanishing. For example, fixing the continuum subtraction points 
the by standard criteria, we have found the following relation for the cut 
Borel transforms of the full and the LC correlators for $m_Q\to\infty$,  
$\mu_B\to\infty$, and $q^2=0$:  
\begin{eqnarray}
\label{result}
\frac{\hat \Gamma_{\rm th}(\mu^2_B\to\infty,q^2=0,\delta_{\rm th})}
{\hat \Gamma_{\rm LC}(\mu^2_B\to\infty,q^2=0,\delta_{\rm LC})}=
1-\frac{4m^2}{3\varepsilon_Q^2 }\log\left(\frac{3\varepsilon_Q}{\sqrt{e}m}\right)+\cdots, 
\end{eqnarray}
the correction being always negative. Here $m$ is the effective constituent mass of the light quark, 
which emerges from the BS equation, $m\simeq \Lambda_{\rm QCD}$, and $\varepsilon_Q$ is the binding energy
of the heavy meson $M_Q=m_Q+\varepsilon_Q$. 
Taking into account that the constituent quark 
mass remains finite in the chiral limit, we come to the following important conclusion: 
{\it In heavy-to-light decays, there exists no rigorous theoretical limit in which 
the cut LC correlator coincides with the cut full correlator}. 

Thus, the off-light-cone
effects in sum rules for heavy-to-light correlators are not negligible and should be
taken into account. 
\item[3.]
We note that the Borel curves for the full and the LC correlators have similar shapes. 
However the light-cone correlator systematically {\it overestimates} the full correlator, the difference at small $q^2$
being $10\div 20$\% in a wide range of the heavy-quark mass relevant for charm and 
beauty decays. We want to point out that the similarity of the Borel 
curves for the full and the LC correlators implies that the systematic difference  
between the correlators cannot be diminished by a relevant choice of the criterion 
for extracting the heavy-to-light form factor. 

The observed effect might suggest a systematic uncertainty in the results for 
form factors obtained within light-cone sum rules. 
As follows from the relation (\ref{result}), this uncertainty is expected to be larger 
for decays of heavy mesons containing the strange quark,
$B_s$ and $D_s$, than for the $B$ and $D$ mesons. This issue deserves further investigation.
\end{itemize}
Finally, we point out the following: Although the model, discussed here, in many aspects  
differs from QCD, this model mimics correctly those  
features which are essential for the effects discussed. 
Therefore, many of the results obtained in this paper are valid also for QCD.  
In particular, the expression (\ref{result}) suggests the following relationship between the light-cone 
and the full correlators in QCD for large values of $m_Q$ and $\mu_B$: 
\begin{eqnarray}
\label{result_QCD}
\frac{\hat \Gamma_{\rm th}(\mu^2_B,q^2,\delta_{\rm th})}{
\hat \Gamma_{\rm LC}(\mu^2_B,q^2,\delta_{\rm LC})}=
1-O\left(\frac{\Lambda_{\rm QCD}}{\delta}\right).
\end{eqnarray}
In numerical estimates, we used the parameters relevant for $B$ and $D$ decays.  
We therefore believe that also the numerical estimates for higher-twist effects 
obtained in this work provide a realistic estimate for higher-twist effects in QCD.   

\vspace{.3cm}
\noindent
{\it Acknowledgments.}
We are grateful to Vladimir Braun, Pietro Colangelo, and Alexander Khodjamirian 
for valuable comments on the preliminary version of the paper.  
We thank Vittorio Lubicz and Matthias Neubert for interesting and stimulating 
discussions. 
D.~M. gratefully acknowledges financial support from INFN, University
``Roma Tre'', 
and the Austrian Science Fund (FWF) under project P17692.

\end{document}